\documentclass[aps,amsmath,amssymb,showpacs,lengthcheck,pre]{revtex4}
\usepackage{graphicx}
\usepackage[english]{babel}
\usepackage{mhchem}
\usepackage{mathtools}
\begin{document}

\newcommand{\be}{\begin{equation}}
\newcommand{\ee}{\end{equation}}
\newcommand{\bea}{\begin{eqnarray}}
\newcommand{\eea}{\end{eqnarray}}
\newcommand{\lan}{\left\langle}
\newcommand{\ran}{\right\rangle}
\newcommand{\ba}{\mathbf{a}}
\newcommand{\bp}{\mathbf{p}}
\newcommand{\bq}{\mathbf{q}}
\newcommand{\bom}{\mathbf{\Omega}}
\newcommand{\bP}{\mathbf{P}}
\newcommand{\bE}{\mathbf{E}}
\newcommand{\bx}{\mathbf{x}}
\newcommand{\bo}{\mathbf{\Omega_d}}
\newcommand{\bk}{\mathbf{k}}
\newcommand{\epa}{\varepsilon_\parallel}
\newcommand{\epe}{\varepsilon_\perp}
\newcommand{\eal}{\varepsilon_\alpha}
\newcommand{\eve}{\varepsilon_v}
\newcommand{\eo}{\varepsilon_0}
\newcommand{\eb}{\varepsilon_b}
\newcommand{\eme}{\varepsilon_m}
\newcommand{\ew}{\varepsilon_w}
\newcommand{\tv}{\tilde{v}}
\newcommand{\tw}{\tilde{w}}
\newcommand{\tU}{\tilde{U}}
\newcommand{\tJv}{\mathbf{\tilde{J}_v}}
\newcommand{\tJ}{\tilde{\mathbf{J}}}
\newcommand{\tphi}{\tilde{\phi}}
\newcommand{\tvarphi}{\tilde{\varphi}}
\newcommand{\tbphi}{\tilde{\mathbf{\varPhi}}}
\newcommand{\tbpsi}{\tilde{\mathbf{\varPsi}}}
\newcommand{\tbV}{\tilde{\mathbf{V}}}
\newcommand{\tbG}{\tilde{\mathbf{G}}}
\newcommand{\tpsi}{\tilde{\psi}}
\newcommand{\tG}{\tilde{G}}
\newcommand{\bz}{\bar{z}}
\newcommand{\bd}{\bar{\Delta}}
\newcommand{\pa}{\parallel}

\title[Beyond Poisson-Boltzmann]{Beyond Poisson-Boltzmann: fluctuations and fluid structure in a self-consistent theory}

\author{S. Buyukdagli$^1$ and R. Blossey$^2$}

\address{$^1$ Department of Physics, Bilkent University, Ankara 06800, Turkey\\
$^2$University of Lille 1, CNRS, UMR 8576 UGSF - Unit\'e de Glycobiologie Structurale et Fonctionnelle, 59000 Lille, France}  
\vspace{10pt}

\begin{abstract}
Poisson-Boltzmann (PB) theory is the classic approach to soft matter electrostatics which has been applied to numerous problems of physical chemistry 
and biophysics. Its essential limitations are the neglect of correlation effects and of fluid structure. Recently, several theoretical insights have allowed
the formulation of approaches that go beyond PB theory in a systematic way. In this topical review we provide an update on the developments achieved 
in self-consistent formulations of correlation-corrected Poisson-Boltzmann theory. We introduce the corresponding system of coupled nonlinear equations 
for both continuum electrostatics with a uniform dielectric constant and a structured solvent, a dipolar Coulomb fluid, including nonlocal effects. While the
approach is only approximate and also limited to corrections in the so-called weak fluctuation regime, it allows to include physically relevant effects, 
as we show for a range of applications of these equations.
\end{abstract}

\maketitle

\section{Introduction}

Poisson-Boltzmann (PB) theory is the cornerstone of soft matter electrostatics, but in recent years several shortcomings of this theory have also been clearly revealed.
PB theory is a mean-field theory, hence it neglects all fluctuation or correlation effects, and as a simple continuum theory it also ignores the structure of solvent 
and ions. In the presence of ions of high valency, prominent in particular in biological systems, the theory fails even qualitatively. A systematic field-theoretic approach 
to soft matter electrostatics developed on the counter-ion case allowed the identification of a coupling parameter \cite{netz00}, 
\begin{equation} 
\label{eq1}
\Xi \equiv \frac{q^3|\sigma| e^4\beta^2}{8\pi \varepsilon^2} = q^2\frac{\ell_B}{\ell_{CG}}
\end{equation} 
where $q$ is the valency of the counter-ions, $\sigma e$ is the surface charge density with electronic charge $e$, $\varepsilon $ is the dielectric constant,
and $\beta = 1/k_B T$. $\Xi$  thus is essentially the ratio of the Bjerrum length $\ell_B = e^2/(4\pi \varepsilon_0 \varepsilon_w k_BT)$ and the 
Gouy-Chapman length $\ell_{GC} = 1/(2\pi \ell_B q |\sigma_s|)$.  
Poisson-Boltzmann theory is the weak coupling limit $\Xi \rightarrow 0$ of the more general theory, while for $\Xi \rightarrow \infty$, the 
strong coupling case, a single-particle picture emerges \cite{netz01}. 

Even within the weak coupling limit, or for intermediate values of the coupling parameter, Poisson-Boltzmann theory does not fully describe 
electrostatic phenomena in soft matter systems. Being a mean-field theory, it entirely lacks correlation effects between the charges. These are,
however, crucial in many physical settings. For the case of electrostatics near macromolecular surfaces or membrane interfaces - one of the most basic situations encountered in
soft matter, this omission does not allow to treat image charge effects of solvated ions. Another crucial effect is the charge reversal of macromolecules induced by
the overscreening of their bare charge by multivalent counterions. This effect can indeed modify the interactions between charged objects even in a qualitative way. 
 
Therefore, in order to remedy this deficit, methods to include fluctuation effects have been devised. In this Topical review, we deal exclusively with
one such approach, which relies on a variational formalism, leading to self-consistently coupled equations of the electrostatic potential and its
correlation function. The formulation that we base ourselves on was originally introduced by Netz and Orland \cite{netz03}, following earlier work 
by Avdeev et al.\cite{avdeev86}. More precisely, Netz and Orland used the variational approach in order to calculate the mean-field level charge renormalisation associated with the 
non-linearity of the Poisson-Boltzmann approach, without considering the correlation effects embodied in the self-consistent equations. 

In recent years, the variational approach has however seen a number of physically relevant applications, covering different charge geometries,
and even dynamical situations such as flow-related effects in nanopores. Hatlo et al. considered the variational formulation of inhomogeneous electrolytes by introducing a restricted self-consistent scheme~\cite{hatlo08}.
In Refs.~\cite{buyuk12,buyuk14I}, one of us (SB) introduced a numerical scheme for the exact solution of the variational equations in slit and cylindrical nanopores. At this point, one should also mention
the one-loop treatment of charge correlations that allows an analytical treatment of inhomogeneous electrolytes. Netz introduced the one-loop calculation of ion partition at membrane surfaces in counterion-only liquids~\cite{netz02}. This was subsequently extended by Lau~\cite{lau08} to electrolytes symmetrically partitioned around a thin charged plane. In Ref.~\cite{buyuk12}, we integrated the one-loop equations of a charge liquid in contact with a thick dielectric membrane.  Finally, in Ref.~\cite{buyuk14V,buyuk15}, we considered the role played by charge correlations on the electrophoretic and pressure-driven DNA translocation through nanopores.

In addition, while the original self-consistent equations have only covered the case of systems that can be described by macroscopic dielectric constants, modified equations have been derived that can also include effects from fluid  structure. The  first dipolar Poisson-Boltzmann theory including solvent molecules as point-dipoles was introduced in Ref.~\cite{coalson96}. Abrashkin et al. incorporated into this model excluded volume effects~\cite{abrashkin07}. One of us (SB) derived a Poisson-Boltzmann equation that relaxes the point-dipole approximation and includes solvent molecules as finite size dipoles~\cite{buyuk13II}.  This model that also accounts for the ionic polarizability was shown to contain the non-locality of electrostatic interactions observed in Molecular Dynamics simulations. Finally, we derived the dipolar self-consistent equations of this model and generalized in this way Netz-Orland's variational equations to explicit solvent liquids~\cite{buyuk14II}.

Our ambition in this topical review is to provide a quick technical introduction to the method and the results that have been achieved with this approach.
We have attempted to make it accessible to a newcomer to the approach by giving sufficient amount of technical detail for the simpler cases. This level
of detail then, by necessity, diminishes for the more complex ones that follow, but we hope that by that time a reader willing to go through about the first
third of the equations in a stepwise manner will have no difficulty in following the rest of the paper. For the latter part, as in all reviews, we refer our readers
to the original articles for further details.

The material presented in the review is organized as follows. In Section 2 we derive the model equations which have been called either variational PB equations, 
self-consistent field equations, or fluctuation-enhanced Poisson-Boltzmann equations (FE-PB), for the case of a system with 1-1 salt. We also discuss the
limits of validity that can be expected from the equations. In Section 3 we review results for systems whose dielectric properties can be properly covered by 
dielectric constants. In particular, we discuss situations of high current interest, the application of the approach to nanopore geometries. Section 4 contains 
very recent extensions of the approach, the fluctuation-enhanced Poisson-Boltzmann equations for a dipolar solvent, the DPBL-equation, as well as
a nonlocal version of the latter. Section 5 presents a brief summary and outlook. We finally note that we explain, wherever possible at present, the theoretical
results closely in relation to experimental findings. This is obviously the ultimate way to validate a theory, and the reader is invited to see how far the 
self-consistent approach to soft matter electrostatics is carrying so far.

\section{The fluctuation-enhanced Poisson-Boltzmann equations} 

\subsection{Derivation.} The fluctuation-enhanced Poisson-Boltzmann equations result from the observation that a simple perturbative treatment of the non-linear
Poisson-Boltzmann equation has only poor convergence properties. This is in particular the case of electrolytes in contact with low permittivity macromolecules or membranes where the singularity of the resulting 
image-charge potential does not allow the one-loop expansion of the grand potential. As in many other branches of physics, variational approaches thus come
as an often fruitful alternative. In this vein, the starting point is the Gibbs variational procedure that consists in minimizing the variational grand potential in the form~\cite{netz03} $ \Omega_v=\Omega_0+\lan H-H_0\ran_0/\Xi $,
where $H_0[\phi]$ is a trial Hamiltonian functional and the bracket $\lan\cdot\ran_0$ denotes the field-theoretic average with respect to this Hamiltonian.
For this Hamiltonian, the most general functional ansatz is a gaussian one including the mean electrostatic potential
$\Phi$ and the covariance of the field expressed via its Green's function $G$ as variational parameters
\begin{equation} \label{eq2}
H_0[\phi] = \frac{1}{2}\int_{\bf r}\int_{\bf r'} \left[\phi({\bf r}) + i\Phi({\bf r}) \right] (\Xi G({\bf r},{\bf r'}))^{-1}\left[\phi({\bf r'}) + i \Phi({\bf r'})\right] \,.
\end{equation} 
For definiteness, we now consider the case of monovalent ions with charges $\pm q$ confined to a region $\Omega$ in presence of a fixed charge density $\varrho_f$.
The Hamiltonian is then, following \cite{xu14}
\begin{equation}  \label{eq3}
H[\phi] = \frac{1}{2\pi}\int_r\left[\frac{(\nabla \phi)^2}{2} + i\varrho_f\phi - \frac{\Lambda}{2} e^{\Xi G_0({\bf r},{\bf r})/2}\cos \phi\right]\,, 
\end{equation}
where $\Lambda$ is the fugacity of the ions, and $G_0({\bf r},{\bf r'}) = 1/|{\bf r} - {\bf r'}|$ is the bare Coulomb potential. The introduction of $G_0$ at this
level takes care of the regularization of the Green's function $G({\bf r},{\bf r'})$ in the final equations as the ionic self-energy corresponding to the equal-point correlation function $G({\bf r},{\bf r})$  diverges in the present dielectric continuum formalism. In eq.({\ref{eq3}), this factor shifts the chemical potential of the ions. With this ansatz, we can compute the grand potential $\Omega$
from which the sought equations follows after extremization with respect to the functions $\Phi$ and $G$. These self-consistent equations read as
\begin{eqnarray} 
&& \nabla^2\Phi({\bf r}) - \Lambda e^{-\Xi c({\bf r})/2}\sinh \Phi({\bf r}) = - 2\varrho_f({\bf r}), \label{eq4}\\
&& \hspace{-6mm} \left[\nabla^2 - \Lambda e^{-\Xi c({\bf r})/2}\cosh \Phi({\bf r})\right]G({\bf r},{\bf r'}) = - 4\pi\delta({\bf r} - {\bf r'}), \label{eq5}\\
&& c({\bf r}) = \lim_{{\bf r} \rightarrow {\bf r'}} \left[G({\bf r},{\bf r'}) - G_0({\bf r},{\bf r'})\right] \label{eq6}
\end{eqnarray} 
Equation (\ref{eq4}) is a modified Poisson-Boltzmann equation, augmented by the correlation function (\ref{eq6}) in the exponential. The correlation function fulfills
a modified Debye-H\"uckel (DH) equation, eq.(\ref{eq5}), in which the usual inverse Debye-length $\kappa^2$ is replaced by a nonlinear function of 
both $c({\bf r})$ and $\phi({\bf r})$. These equations are one realization (for the case of 1-1 salt, and without further specification of the fixed charge geometries) 
of the self-consistent or fluctuation-enhanced Poisson-Boltzmann equations. One should also note that Eqs.(4)-(5) can be easily generalized to an asymmetrical
electrolyte (see e.g. Ref. [7]).

\subsection{Validity.} The self-consistent equations (\ref{eq4}) - (\ref{eq6}) are, by their very construction, only approximate. Their validity ultimately rests on the
validity of the Gaussian assumption to begin with, and this is, as usual in variational approaches, not always easy to quantify.  One can, however, identify qualitatively the validity regime by considering charge correlations in a bulk electrolyte. In this case, the electrostatic potential $\phi$ vanishes and we are left with a Debye-H\"uckel type equation for the Green's function \cite{pujos14}
\begin{equation} \label{eq7}
-\nabla^2G({\bf r},{\bf r'}) + \Lambda e^{-\Xi c(\bf{r})/2} G({\bf r},{\bf r'}) = 4\pi\delta({\bf r} - {\bf r'})\, .
\end{equation}
If we define the screening parameter as $ \kappa^2 \equiv \Lambda e^{-\Xi c(\bf{r})/2}$, the Green's function becomes
\begin{equation} \label{eq8} 
G({\bf r},{\bf r'}) = \frac{e^{-\kappa|{\bf r} - {\bf r'}|}}{|{\bf r} - {\bf r'}|}.
\end{equation}
Inserting Eq.~(\ref{eq8}) into Eq.(\ref{eq6}), one finds $c = - \kappa$. We are thus led to a first self-consistency condition given by
\begin{equation} \label{eq9}
\kappa^2 = \Lambda e^{\Xi \kappa/2}\, . 
\end{equation}
This equation ceases to have a solution for large values of the coupling parameter $\Xi$, clearly indicating that the self-consistent equations~(\ref{eq5})-(\ref{eq6}) are only valid for weak to moderate charge correlations. This condition can be, in turn, quantified through Eq.~(\ref{eq1}) in terms of the model parameters.

In inhomogeneous liquids, the validity limit of the self-consistent equations~(\ref{eq5})-(\ref{eq6}) is not so easy to assess, as they are, due to their highly non-linear character,
not amenable to analytic solutions, even for simple geometries. Numerical methods have been recently developed to solve them \cite{buyuk12,buyuk14I,xu14,deng14}.
In particular, by comparison with MC simulations, Refs.~\cite{buyuk12} and~\cite{buyuk14I} identified the validity regime of the equations for liquids confined to slit and cylindrical nanopores, respectively. In the following we will discuss approximate solutions of Eqs.~(\ref{eq4})-(\ref{eq6}) and their modifications to physically relevant situations which to some extent also permit analytical calculations. In particular, we confront these solutions to data from experiment and simulations.

\section{The variational equations for a dielectric continuum}

In this section we turn to the application of the SC equations to some specific physical situations. First, we discuss ion correlations and charge reversal at a planar interface,
and then move on to dynamic effects associated with DNA translocation through membrane pores. These examples were selected as they allow to convey the main
insights on correlation effectsthat can be gained from the variational approach, and in conjunction with ions of higher valency, where PB-theory is known to fail. 
Further applications of the equations in the case of a dielectric continuum concerned the effect of image charges on macro-ions \cite{lee09} and on the electrical double layer  \cite{wang13,wang15}. Ion size effects upon ionic exclusion from dielectric interfaces and slit nanopores were treated in \cite{buyuk10,buyuk11I,buyuk11II}. The modification of ion polarizabilities from the gas phase to solvation in polar liquids were discussed in \cite{buyuk13}. 

\subsection{One-loop expansion of SC equations and charge reversal}

We begin the discussion with the one-loop (1-$\ell$) expansion of the electrostatic SC equations valid exclusively for dielectrically continuous systems 
$\varepsilon(\mathbf{r})=\varepsilon_w$. We consider a symmetric electrolyte composed of two ionic species with valencies $\pm q$ and bulk density 
$\rho_b$. In order to facilitate the link to the research literature, in passing from Eq.~(\ref{eq4}) to the subsequent Eq.~(\ref{eq10}) we will introduce 
the definition of the new average potential $\psi(\mathbf{r})=-q\Phi(\mathbf{r})$ and the Green's function $v(\mathbf{r},\mathbf{r}')=\Xi G(\mathbf{r},\mathbf{r}')$.  
For this case, the SC equations read as
\begin{equation} \label{eq10}
\nabla^2\psi(\mathbf{r})-\kappa_b^2e^{-V_w(\mathbf{r})-\frac{q^2}{2}\delta v(\mathbf{r},\mathbf{r})}\sinh\left[\psi(\mathbf{r})\right] = \\
-4\pi\ell_Bq\sigma(\mathbf{r})
\end{equation}
\begin{eqnarray}
\label{eq11} 
\left\{\nabla^2 -\kappa_b^2e^{-V_w(\mathbf{r} )-\frac{q^2}{2}\delta v(\mathbf{r},\mathbf{r})}\cosh\left[\psi(\mathbf{r})\right]\right\}v(\mathbf{r},\mathbf{r}') = \nonumber \\ 
-4\pi\ell_B\delta(\mathbf{r}-\mathbf{r}'), 
\end{eqnarray}
where the ionic self-energy (or renormalized equal-point correlation function) is defined by
\be\label{eq12}
\delta v(\mathbf{r}) \equiv \delta v(\mathbf{r},\mathbf{r})=\ell_B\kappa_b+ v(\mathbf{r},\mathbf{r})-v_c^b(0),
\ee
with the DH screening parameter $\kappa_b^2=8\pi\ell_Bq^2$, and the function $V_w(\mathbf{r})$ is the ionic steric potential accounting for the rigid boundaries in the system. 

The 1-$\ell$ expansion of these equations consists in Taylor-expanding Eqs.~(\ref{eq10})-(\ref{eq11}) in terms of the electrostatic Green's function $v(\mathbf{r},\mathbf{r}')$. 
Splitting the average potential into its MF and 1-$\ell$ components as $\psi(\mathbf{r})=\psi_0(\mathbf{r})+\psi_1(\mathbf{r})$, one obtains for the MF potential and the 
1-$\ell$ electrostatic Green's function the equations
\bea\label{eq13}
\nabla^2\psi_0(\mathbf{r})-\kappa_b^2e^{-V_w(\mathbf{r})}\sinh\left[\psi_0(\mathbf{r})\right]=-4\pi\ell_B\sigma(\mathbf{r}),
\eea
\bea
\label{eq14}
\left\{\nabla^2-\kappa_b^2e^{-V_w(\mathbf{r})}\cosh\left[\psi_0(\mathbf{r})\right]\right\}v(\mathbf{r},\mathbf{r}') = \nonumber \\
-4\pi\ell_B\delta(\mathbf{r}-\mathbf{r}'),
\eea
and for the 1-$\ell$ correction to the average potential 
\bea
\label{eq15}
\left\{\nabla^2-\kappa_b^2e^{-V_w(\mathbf{r})}\cosh\left[\psi_0(\mathbf{r})\right]\right\}\psi_1(\mathbf{r}) = \nonumber \\
-\frac{q^2}{2}\kappa_b^2e^{-V_w(\mathbf{r})}\delta v(\mathbf{r})\sinh\left[\psi_0(\mathbf{r})\right].
\eea
In order to solve the system of differential equations~(\ref{eq13})-(\ref{eq15}), we first invert Eq.~(\ref{eq14}) and recast it in a more practical form for analytical evaluation. 
By using the definition of the Green's function
\be\label{eq16}
\int\mathrm{d}\mathbf{r}_1v^{-1}(\mathbf{r},\mathbf{r}_1)v(\mathbf{r}_1,\mathbf{r}')=\delta(\mathbf{r}-\mathbf{r}'),
\ee
one can invert Eq.~(\ref{eq14}) and express the electrostatic kernel as
\be\label{eq17}
v^{-1}(\mathbf{r},\mathbf{r}')=-\frac{1}{4\pi\ell_B}\left\{\nabla_\mathbf{r}^2-\kappa_b^2e^{-V_w(\mathbf{r})}\cosh\left[\psi_0(\mathbf{r})\right]\right\}\delta(\mathbf{r}-\mathbf{r}'),
\ee
in terms of which Eq.~(\ref{eq15}) can be written as
\be
\label{eq18}
\int\mathrm{d}\mathbf{r}_1v^{-1}(\mathbf{r}_2,\mathbf{r}_1)\psi_1(\mathbf{r}_1)=q^4\rho_be^{-V_w(\mathbf{r}_2)}\delta v(\mathbf{r}_2)\sinh\left[\psi_0(\mathbf{r}_2)\right].
\ee
Multiplying Eq.~(\ref{eq18}) with the potential $v(\mathbf{r},\mathbf{r}_2)$ and integrating over the variable $\mathbf{r}_2$, one finally gets the integral relation for the 
1-$\ell$ potential correction completing the equations~(\ref{eq13})-(\ref{eq14})
\be\label{eq19}
\psi_1(\mathbf{r})=\rho_bq^4\int\mathrm{d}\mathbf{r}_1e^{-V_w(\mathbf{r}_1)}v(\mathbf{r},\mathbf{r}_1)\delta v(\mathbf{r}_1)\sinh\left[\psi_0(\mathbf{r}_1)\right].
\ee
We also note that within 1-$\ell$-theory the ion number densities are given by
\be\label{eq20}
\rho_{\pm}(\mathbf{r})=\rho_be^{-V_w(\mathbf{r})\mp\psi_0(\mathbf{r})}\left\{1-\frac{q^2}{2}\delta v(\mathbf{r})\mp\psi_1(\mathbf{r})\right\}.
\ee
We now have the main set of equations ready and turn to the analytical solutions of Eqs.~(\ref{eq13})-(\ref{eq15}) for a simple planar geometry in order to investigate the charge 
inversion phenomenon~\cite{buyuk14I}. Then we will couple these equations with the Stokes equation and show that charge inversion gives rise to the reversal of the electrophoretic DNA mobility~\cite{buyuk14V} and the hydrodynamically induced ion currents through cylindrical pores~\cite{buyuk15}.

\subsubsection{Ionic correlations and charge reversal at planar interfaces.}

In this part we consider the charge reversal effect in the case of a negatively charged planar interface located at $z=0$. The electrolyte occupies the half-space $z>0$ while 
the left half-space at $z<0$ is ion-free. In the SC-equations this corresponds to a steric potential $V_w(\mathbf{r})=\infty$ if $z<0$ and $V_w(\mathbf{r})=0$ for $z>0$. 
The wall charge distribution function is $\sigma(\mathbf{r})=-\sigma_s\delta(z)$. By introducing the Gouy-Chapman length $\ell_{GC} \equiv \mu=1/(2\pi q\ell_B\sigma_s)$ that corresponds to the thickness of the interfacial counterion layer, the solution of Eq.~(\ref{eq13}) satisfying Gauss' law $\psi'_0(z=0)=2/\mu$ reads as
\be\label{eq21}
\psi_0(z)=-2\ln\left[\frac{1+e^{-\kappa_b(z+z_0)}}{1-e^{-\kappa_b(z+z_0)}}\right],
\ee
where we introduced the parameter $s=\kappa_b\mu$ and the auxiliary functions $z_0=-\ln\left[\gamma(s)\right]/\kappa_b$ and $\gamma(s)=\sqrt{s^2+1}-1$. In the DH-limit of weak surface charge or strong salt, the potential~(\ref{eq21}) becomes $\psi_0(z)\simeq(2/s)e^{-\kappa_bz}$. The parameter $s$ thus corresponds to the inverse magnitude of the surface potential. 

Taking advantage of the planar symmetry in the $(x,y)$-plane one can expand the Green's function in the Fourier basis as
\be\label{eq22}
v(\mathbf{r},\mathbf{r}')=\int_0^\infty\frac{\mathrm{d}kk}{2\pi}\mathrm{J}_0\left(k|\mathbf{r}_\|-\mathbf{r}'_\||\right)\tv_0(z,z'),
\ee
where $\mathbf{r}_\pa$ is the position vector in the $(x,y)$-plane and $J_0$ the Bessel function of order zero. Substituting the expansion (22) together with the 
MF-potential~(\ref{eq21}) into the kernel equation~(\ref{eq11}), the latter takes the form
\bea\label{eq23}
\partial_z^2\left[1-\theta(z)\left\{p_b^2+2\kappa_b^2\mathrm{csch}^2\left[\kappa_b(z+z_0)\right]\right\}\right]\tv(z,z',k) = \nonumber 
\eea 
\be
 -4\pi\ell_B\delta(z-z')
\ee
where we defined the function $p_b=\sqrt{k^2+\kappa_b^2}$. The solution of Eq.~(\ref{eq23}) satisfying the continuity of the potential $\tv(z,z')$ and the displacement field
$\partial_z\tv(z,z')$ at $z=0$ and $z=z'$ reads
\be\label{eq24}
\tv_0(z,z')=\frac{2\pi\ell_B p_b}{k^2}\left[h_+(z)h_-(z')+\Delta h_-(z)h_-(z')\right]
\ee
for $0\leq z\leq z'$, where the homogeneous solutions of Eq.~(\ref{eq23}) are given by
\be\label{eq25}
h_\pm(z)=e^{\pm p_bz}\left\{1\mp\frac{\kappa_b}{p_b}\coth\left[\kappa_b(z+z_0)\right]\right\},
\ee
and the $\Delta$-function reads
\be\label{eq26}
\Delta=\frac{\kappa_b^2\mathrm{csch}^2\left(\kappa_bz_0\right)+(p_b-k)\left[p_b-\kappa_b\coth\left(\kappa_bz_0\right)\right]}
{\kappa_b^2\mathrm{csch}^2\left(\kappa_bz_0\right)+(p_b+k)\left[p_b+\kappa_b\coth\left(\kappa_bz_0\right)\right]}\, .
\ee
For $z\geq z'$, the solution of Eq.~(\ref{eq23}) can be obtained by interchanging the variables $z$ and $z'$ in Eq.~(\ref{eq24}).
\begin{figure}
\begin{center}
(a)\includegraphics[width=0.75\linewidth]{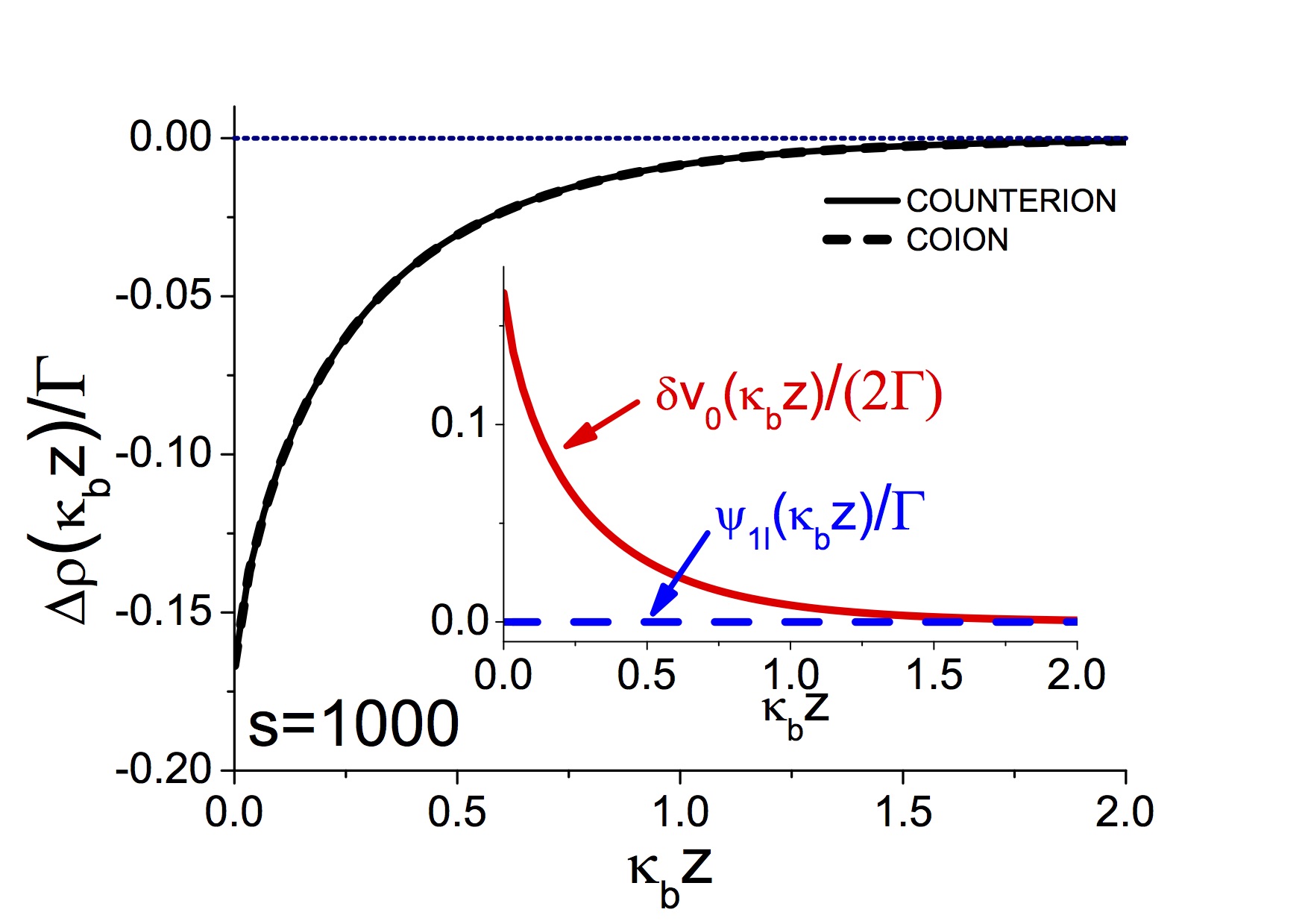}
(b)\includegraphics[width=0.75\linewidth]{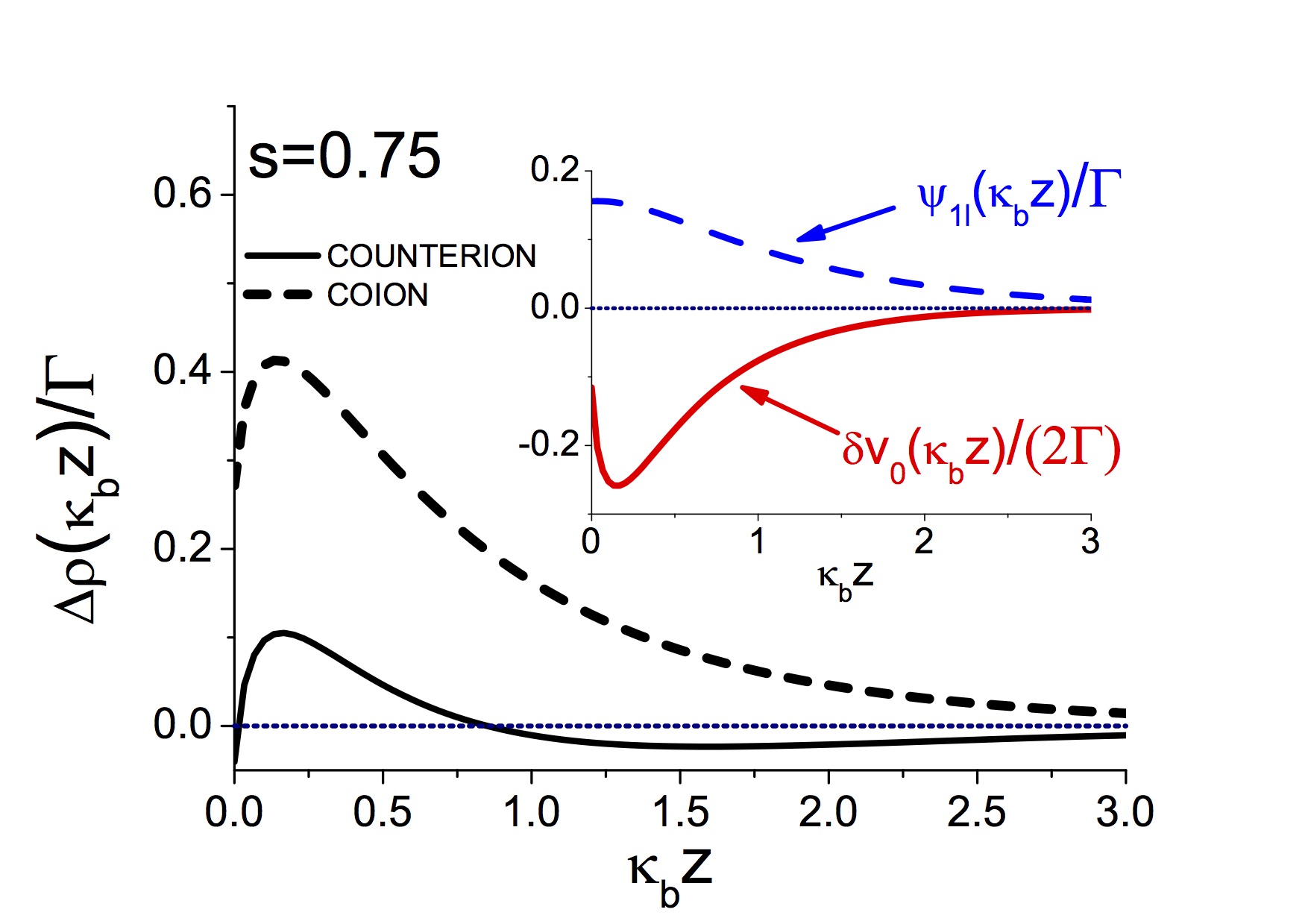}
\caption{(Color online) One-loop corrections to the counter-ion (solid lines) and co-ion densities (dashed lines) from Eq.~(\ref{eq31}),
for two different values of the parameter $s = \kappa_b\mu$, the ratio of the Gouy-Chapman and the screening length.
Case (a): $s=1000$ (a weakly charged membrane) and case (b): $s=0.75$ (a strongly charged membrane). The inset shows the ionic self-energy 
and the one-loop correction to the external potential for these parameters.}
\end{center}
\label{fig1}
\end{figure}
From now on we switch to the non-dimensionalized coordinate $\bz=\kappa_bz$. Rescaling as well the wave-vector of the Fourier expansion (22) 
as $k\to u=p_b/\kappa_b$ and inserting the function~(\ref{eq24}) into Eq.~(\ref{eq12}), the ionic self-energy accounting for charge correlations takes the form
\bea\label{eq27}
\delta v_0(\bz)&=&\Gamma\int_1^\infty\frac{\mathrm{d}u}{u^2-1}\left\{-\mathrm{csch}^2\left[\bz-\ln\gamma_c(s)\right]\right.\\
&&\left.+\bd\left(u+\coth\left[\bz-\ln\gamma_c(s)\right]\right)^2e^{-2\bz u}\right\},\nonumber
\eea
where we defined the electrostatic coupling parameter $\Gamma=\ell_B\kappa_b$ and
\be\label{eq28}
\bd=\frac{1+s\left(su-\sqrt{s^2+1}\right)\left(u-\sqrt{u^2-1}\right)}{1+s\left(su+\sqrt{s^2+1}\right)\left(u+\sqrt{u^2-1}\right)}.
\ee
Inserting into Eq.~(\ref{eq19}) the MF-potential~(\ref{eq21}), the Green's function~(\ref{eq24}), the self-energy~(\ref{eq27}), and carrying out the spatial integral, 
the 1-$\ell$-correction to the average potential takes the form
\be\label{eq29}
\psi_{1\ell}(\bz)=\frac{q^2}{4}\Gamma\mathrm{csch}\left[\bar{z}-\ln\gamma_c(s)\right]\int_1^\infty\frac{\mathrm{d}u}{u^2-1}F(\bz,u),
\ee
where we introduced the auxiliary function
\bea\label{eq30}
F(\bz,u)&=&\frac{2+s^2}{s\sqrt{1+s^2}}-\bd\left(\frac{1}{u}+2u+\frac{2+3s^2}{s\sqrt{1+s^2}}\right)\nonumber \\ 
&&+\frac{\bd}{u}e^{-2u\bz}+\left(\bd e^{-2u\bz}-1\right)\coth\left[\bz-\ln\gamma_c(s)\right]. \\ 
&& \nonumber
\eea
We note that the correlation-corrected ion densities~(\ref{eq20}) are fully characterized by the potentials~(\ref{eq21}), (\ref{eq27}), and~(\ref{eq29}). One also sees that these 
functions depend solely on the parameters $s$ and $\Gamma$.

In Figure 1, we show the 1-$\ell$-correction to the ion densities
\be\label{eq31}
\Delta\rho_\pm(\bz) \equiv \frac{\rho_\pm(\bz)-\rho_\pm^{MF}(\bz)}{\rho_\pm^{MF}(\bz)}=-\frac{q^2}{2}\delta v_0(\bz)\mp\psi_{1\ell}(\bz),
\ee
where the MF-ion densities are given by $\rho_\pm^{MF}(\bz)=\rho_be^{\mp\psi_0(\bz)}$. The top plot (a) illustrates the density correction at a weakly charged membrane ($s=1000$).  In this case, the absence of ions in the left half-space results in the ionic screening deficiency close to the membrane surface. As a result, ions prefer to move away from the interface towards the bulk where they are more efficiently screened and possess a lower free energy. This translates in turn into a positive ionic self-energy $\delta v_0(\bz)>0$ (see the inset) and a decrease of the MF-level co-ion and counter-ion densities (main plot) by charge correlations. 

In the plot of Figure 1 (b), we consider a strongly charged membrane ($s=0.75$). In this parameter regime, the strong counterion attraction results in an interfacial charge excess. Ions being more efficiently screened in the vicinity of the membrane surface, they tend to approach the interface. This effect is reflected in the attractive ionic self-energy  $\delta v_0(\bz)<0$ (inset) and the amplification of the MF-density of both coions and counterions (main plot). In Ref.~\cite{buyuk12}, it was shown that the transition between these two regimes with the surface self-energy $\delta v_0(0)$ switching from positive to negative occurs when the size of the interfacial counterion layer becomes comparable to the ionic screening radius in the bulk region, i.e. at $\mu\simeq\kappa_b^{-1}$. This equality yields the characteristic membrane charge $\sigma_s^*=\sqrt{2\rho_b/(\pi\ell_B)}$ above which the interfacial screening dominates the bulk screening.
\begin{figure}
\begin{center}
\includegraphics[width=0.8\linewidth]{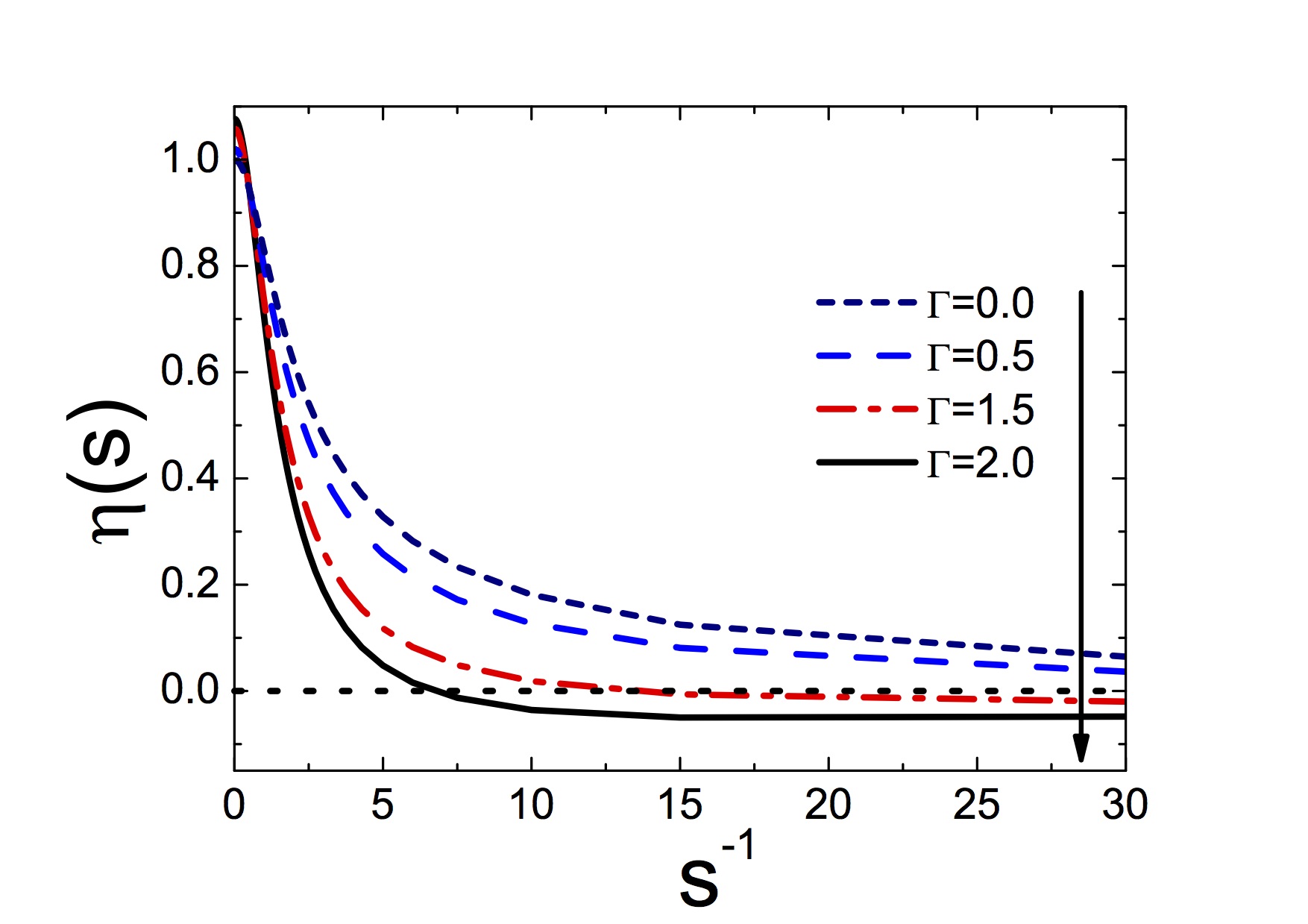}
\caption{(Color online) Charge renormalization factor $\eta(s)$, see Eq. (\ref{eq33}) against $s^{-1}$, $s = \kappa_b\mu$,
for several values of the coupling parameter $\Gamma = \ell_B \kappa_b $.}
\end{center}
\label{fig2}
\end{figure}

For a negatively charged membrane, the MF-potential is negative (see Eq.~(\ref{eq18})). Furthermore, in the inset of Figure 1(b) where we plotted the 1-$\ell$-correction 
Eq.~(\ref{eq29}) to the average potential, one sees that the former is positive. This stems from the interfacial screening excess. Thus, at strongly charged membranes, correlations attenuate the magnitude of the negative MF-potential. 

We will now show that this peculiarity is the precursor of the \textit{charge reversal} effect. In order to illustrate this point, we note that  at large distances from the charged plane $\bz\gg1$, the MF-potential~(\ref{eq21}) behaves as $\psi_0(\bz)\simeq-4\gamma(s)e^{-\bz}$. Expanding the 1-$\ell$-correction to the average electrostatic potential~(\ref{eq30}) in the 
same limit, one finds that the correlation-corrected average potential $\psi_{1\ell}(\bz)=\psi_0(\bz)+\psi_1(\bz)$ takes the form
\be\label{eq32}
\psi_{1l}(\bz)\simeq-\frac{2}{s}\eta(s)e^{-\bz},
\ee
where we introduced the charge renormalization factor accounting for electrostatic many-body effects
\be\label{eq33}
\eta(s)=2s\gamma(s)\left[1-\frac{q^2\Gamma}{8}\mathrm{I}(s)\right],
\ee
with the auxiliary function
\bea\label{eq34}
\mathrm{I}(s)= 
\eea
\bea
\int_1^\infty\frac{\mathrm{du}}{u^2-1}\left\{\frac{2+s^2}{s\sqrt{1+s^2}}-1 -\bd\left(\frac{1}{u}+2u+\frac{2+3s^2}{s\sqrt{1+s^2}}\right)\right\}. \nonumber
\eea
In Figure 2, we plot Eq.~(\ref{eq33}) versus $s^{-1}$ for various coupling parameters $\Gamma$. In the MF-limit $\Gamma=0$, and the charge renormalization factor $\eta(s)$ accounts for non-linearities neglected by the linearized PB-theory. As the latter overestimates the actual value of the electrostatic potential at strong charges, with increasing surface charge from left to right, the correction factor $\eta(s)$ drops from unity to zero. Moreover, one sees that at the coupling parameter $\Gamma=2$, the charge renormalisation factor passes from positive to negative. In other words, at large distances from the interface, the total average potential~(\ref{eq32}) switches from negative to positive. This is the signature of the \textit{charge inversion} effect. As the adsorbed counterions overcompensate the negative fixed surface charge, the interface acquires an effective positive charge. For small $s$ (or large surface charge density), Eq.~(\ref{eq33}) takes the asymptotic form
\be\label{eq35}
\eta(s)=2s\left[1-\Gamma\frac{\pi-4\ln(2s)}{16}\right]+O\left(s^2\right).
\ee
Setting the equality~(\ref{eq35}) to zero, one finds that the charge reversal takes place at the particular value 
\be\label{eq36}
s_c(\Gamma)=\frac{1}{2}\exp\left(\frac{\pi}{4}-\frac{4}{\Gamma}\right).
\ee
Having established the charge reversal effect in the planar geometry, we will now turn to its influence on ion currents and polymer mobilities through cylindrical nanopores,
which is a problem of current interest in the field of soft matter physics.

\subsubsection{Electrophoretic DNA mobility reversal by multivalent counterions.}

In this part we discuss the effect of charge inversion induced by multivalent ions on the electrophoretic mobility of a DNA molecule~\cite{buyuk14V}. 
The molecule is modeled as a charged cylinder with radius $a=1$ nm, translocating through a nanopore of cylindrical geometry with radius $d=3$ nm. 
The configuration is depicted in Figure 3. The solution of the 1-$\ell$ Eqs.~(\ref{eq13})-(\ref{eq15}) in a cylindrical geometry is similar 
to their solution in planar geometry. Thus, we will skip the technical details here and refer to Refs.~\cite{buyuk14I} and~\cite{buyuk14V} for details. 
The DNA molecule has a negative surface charge distribution $\sigma(r)=-\sigma_p\delta(r-a)$, with $\sigma_p=0.4$ $e/\mathrm{nm}^2$ and $r$ the 
radial distance of ions from the symmetry axis of the cylindrical polymer. In the general case, the permittivity of the nanopore $\varepsilon_m$ and DNA may differ 
from the water permittivity $\varepsilon_w$. However, in order to simplify the technical task, we assume that there is no dielectric discontinuity in the system and set 
$\varepsilon_m=\varepsilon_p=\varepsilon_w=80$. 

As the electrophoretic translocation of DNA under an external potential gradient $\Delta V$  corresponds to the collective motion of the electrolyte and the DNA molecule, we need to derive first the convective fluid velocity $u_c(r)$ given by the Stokes equation
\be\label{eq37}
\eta\nabla^2 u_c(r)+e\rho_c(r)\frac{\Delta V}{L}=0,
\ee
with the viscosity coefficient of water $\eta=8.91\times 10^{-4}\;\mathrm{Pa} \cdot \mathrm{s}$, the nanopore length $L$, and the liquid charge density 
\be\label{eq38}
\rho_c(r)=\sum_iq_i\rho_i(r), 
\ee
where $\rho_i(r)$ is the number density of the ionic species $i$ with valency $q_i$, 
\be\label{eq39}
\rho_i(r)=\rho_{ib}e^{-q_i\psi_0(r)}\left[1-q_i\psi_1(r)-\frac{q_i^2}{2} \delta v(r)\right].
\ee
By making use of the Poisson equation $\nabla^2\psi_{1\ell}(r)+4\pi\ell_B\rho_c(r)=0$ in Eq.~(\ref{eq37}), the latter can be written 
explicitly in the cylindrical geometry as
\be\label{eq40}
\frac{\eta}{r}\frac{\partial}{\partial r}r\frac{\partial}{\partial r}u_c(r)-\frac{k_BT}{er}\frac{\Delta V}{L}\frac{\partial}{\partial r}r\varepsilon(r)\frac{\partial}{\partial r}\psi_{1\ell}(r)=0.
\ee
In order to solve Eq.~(\ref{eq40}), we first note that the 1-$\ell$-potential $\psi_{1\ell}(r)$ satisfies Gauss' law $\phi'(a)=4\pi\ell_B\sigma_p$ at the DNA surface. In the steady-state regime where DNA translocates at constant velocity $v$, the longitudinal electric force per polymer length $F_e=-2\pi ae\sigma_p\Delta V/L$ will compensate the viscous friction force 
$F_v=2\pi a\eta u_c'(a)$, that is $F_e+F_v=0$. Finally, we impose the boundary conditions $u_c(d)=0$ (no-slip) and $u_c(a)=v$. Integrating Eq.~(\ref{eq40}) and imposing the above-mentioned conditions, we get the convective flow velocity 
\be\label{eq41}
u_c(r)=-\mu_e\frac{\Delta V}{L}\left[\phi(d)-\phi(r)\right],
\ee
and the DNA translocation velocity
\be\label{eq42}
v=-\mu_e\frac{\Delta V}{L}\left[\phi(d)-\phi(a)\right],
\ee
with the reduced electrophoretic mobility
\be\label{eq43}
\mu_e=\frac{k_BT\varepsilon_w}{e\eta }.
\ee
\begin{figure}
\begin{center}
\includegraphics[width=0.9\linewidth]{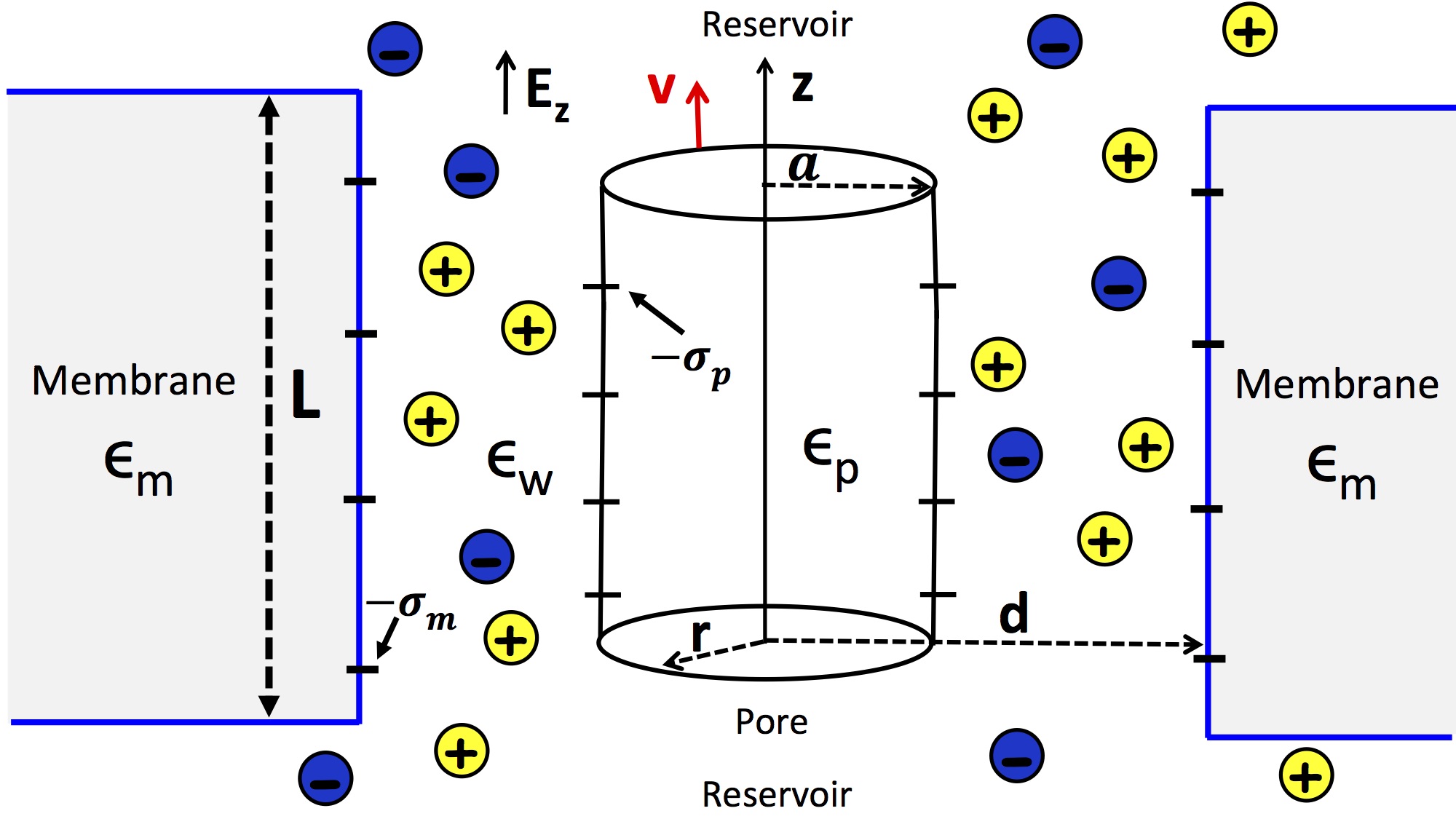}
\caption{(Color online) Nanopore geometry: A cylindrical polyelectrolyte of radius $a=1$ nm, surface charge $-\sigma_p$, and dielectric permittivity $\varepsilon_p$ is confined 
to a cylindrical pore of radius $d=3$ nm, with wall charge $-\sigma_m$, and membrane permittivity $\varepsilon_m$. The permittivity of the electrolyte is $\varepsilon_w=80$.}
\end{center}
\label{Figure3}
\end{figure}
\begin{figure}
\begin{center}
\includegraphics[width=0.9\linewidth]{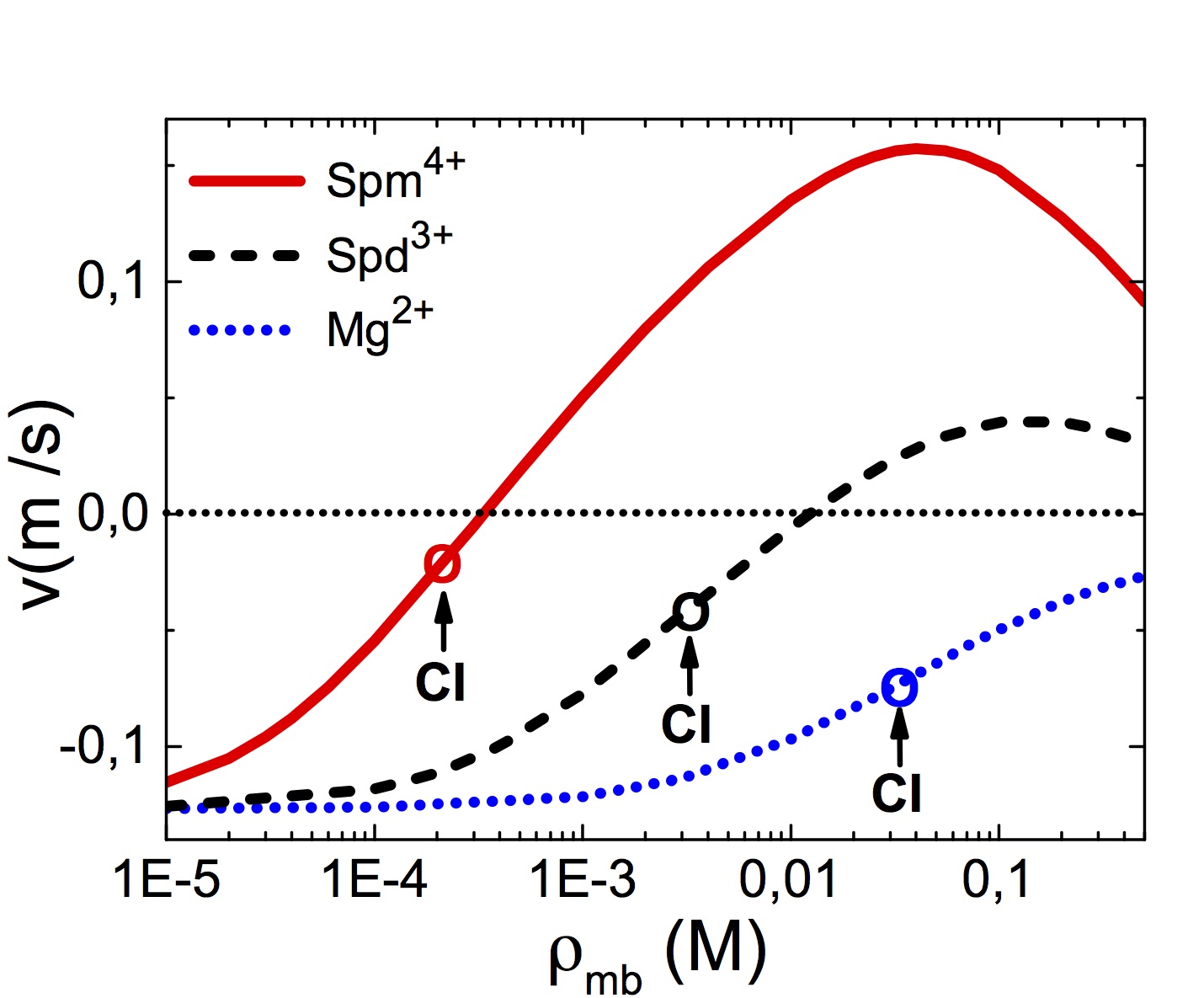}
\caption{(Color online)  Polymer translocation velocity against the density of the multivalent counterion species $\mbox{I}^{m+}$  (see the legend) in the electrolyte mixture KCl+I$\mbox{Cl}_m$. $\mbox{K}^+$ density is fixed at $\rho_{+b}=0.1$ M. The confined double-stranded DNA has surface charge $\sigma_p=0.4$ $\mathrm{e/nm}^2$.  The potential gradient is $\Delta V=120$ mV.  Circles mark the charge inversion (CI) point of the DNA molecule.}
\end{center}
\label{Figure4}
\end{figure}
We now consider an electrolyte mixture of composition KCl+I$\mbox{Cl}_m$ including an arbitrary type of multivalent counterions $\mbox{I}^{m+}$.  The electrophoretic DNA velocity~(\ref{eq42}) against the multivalent ion density $\rho_{mb}$ is displayed in Figure 4 for various multivalent ion types. With divalent $\mbox{Mg}^{2+}$ ions, the DNA translocation velocity is negative. This corresponds qualitatively to the classical MF-transport where the negatively charged DNA molecule moves oppositely to the applied electric field. However, in the mixed electrolytes containing trivalent spermidine and quadrivalent spermine ions, and with the increase of the multivalent ion density, the DNA velocity switches from negative to positive. In other words, at large multivalent counter-ion concentrations, the molecule translocates parallel with the field. Finally, one notes that beyond the mobility reversal density, the positive DNA velocity in spermidine and spermine liquids reaches a peak and drops beyond this point. 
\begin{figure}
\begin{center}
\includegraphics[width=0.9\linewidth]{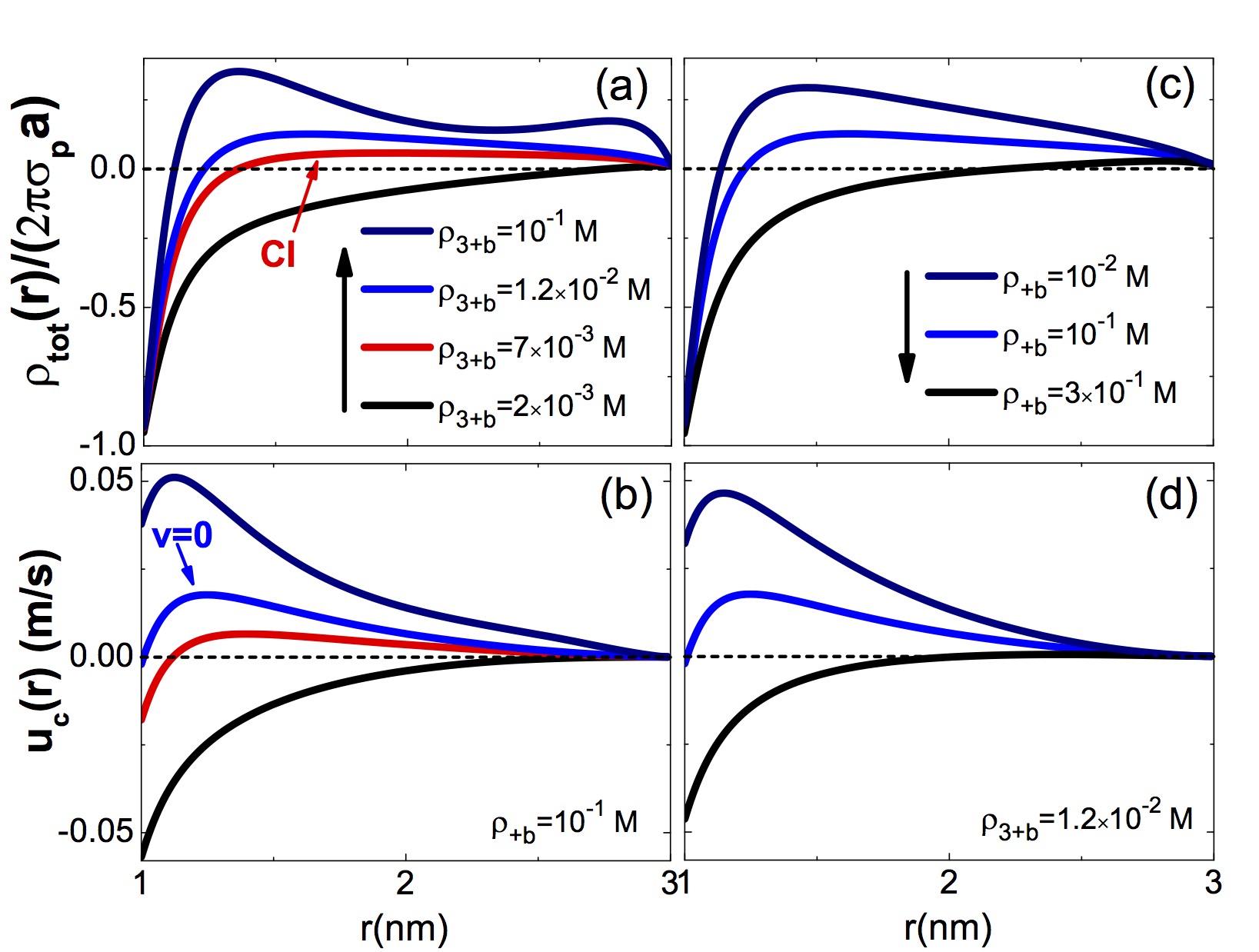}
\caption{(Color online) Cumulative charge densities rescaled with the bare DNA charge (top plots) and solvent velocities (bottom plots) at fixed $\mbox{K}^+$ and varying $\mbox{Spd}^{3+}$ concentration in (a) and (b), and fixed $\mbox{Spd}^{3+}$ and varying  $\mbox{K}^+$ concentration in (c) and (d). In each column, the same colour in the top and bottom plots corresponds to a given bulk counterion  concentration displayed in the legend. The remaining parameters are the same as in Figure 4.}
\label{Figure5}
\end{center}
\end{figure}

In order to explain the reversal of the DNA mobility by multivalent counterions, we plot in Figure 5 (a) the cumulative charge density of the KCl+$\mbox{Spm}^{3+}\mbox{Cl}_3$ 
fluid including the polyelectrolyte charge
\be\label{eq44}
\rho_{tot}(r)=2\pi\int_a^r\mathrm{d}r'r'\left[\rho_c(r')+\sigma_s(r')\right]
\ee
for various bulk $\mbox{Spd}^{3+}$ concentrations. We also illustrate in Figure 5 (b) the cumulative liquid velocity~(\ref{eq42}) at the same densities. At the low $\mbox{Spd}^{3+}$ density $\rho_{3+b}=2\times10^{-3}$ M, moving from the DNA surface towards the pore wall, the total charge density rises from the net DNA charge towards zero. This corresponds qualitatively to the MF-picture where counterions gradually screen the DNA charge. Hence,  the negatively charged fluid and DNA move oppositely to the field, that is $u_c(r)<0$ and $v=u_c(a)<0$.  Increasing now the $\mbox{Spd}^{3+}$ density to $\rho_{3+b}=7\times10^{-3}$ M, the cumulative charge density switches from negative to positive at $r\gtrsim 1.3$ nm. This is the signature of charge reversal where due to correlation effects induced by $\mbox{Spd}^{3+}$ ions, counterions locally overcompensate the DNA charge. Consequently, the liquid flows parallel with the field ($u_c(r)>0$) in the region $r>1.3$ nm. However, because the hydrodynamic drag force is not sufficiently strong to dominate the electrostatic coupling between the DNA molecule and the external electric field, the molecule and the liquid around continue to move in the direction of the field. At the larger $\mbox{Spd}^{3+}$ density $\rho_{3+b}=1.2\times10^{-2}$ M where charge reversal becomes more pronounced, the drag force compensates exactly the electric force on the DNA molecule. As a result, DNA stops its translocation, i.e. $v=u_c(a)=0$. At larger $\mbox{Spd}^{3+}$ concentrations, the DNA molecule and the electrolyte move parallel with the field. We emphasize that an important challenge in DNA translocation consists in the minimization of the DNA translation velocity for an accurate sequencing of its genetic code~\cite{clarke09}. The present result suggests 
that this can be achieved by tuning trivalent or quadrivalent ion densities in the liquid.

To summarize, it is found that the inversion of the DNA mobility is induced by charge reversal. However, charge inversion should also be strong enough for mobility reversal to occur. This can also be seen in Figure 4 where we display the charge inversion points by open circles. One notes that charge inversion precedes the mobility reversal that occurs only with trivalent and quadrivalent ions. We now focus on the peak of the velocity curves in this figure. One notes that at the largest  $\mbox{Spd}^{3+}$ concentration in Figure 5 (a), the first positive peak of the cumulant charge density is followed by a well. The corresponding local drop of the cumulative charge density stems from the attraction of $\mbox{Cl}^-$ ions towards the charge inverted DNA. At higher $\mbox{Spd}^{3+}$ concentrations, the stronger chloride attraction attenuates the charge reversal effect responsible for the mobility inversion. This explains the reduction of the inverted charge mobility at large multivalent densities in Figure 4. Finally, we consider the effect of potassium concentration that can be easily tuned in translocation experiments. In Figures 5 (c) and (d), we show the charge density and velocity at fixed $\mbox{Spd}^{3+}$ concentration for various $\mbox{K}^{+}$ densities. Starting at the charge inverted density values $\rho_{3+b}=1.2\times10^{-2}$ M and $\rho_{+b}=10^{-2}$ M, and raising the bulk potassium concentration from top to bottom, the cumulative charge density is seen to switch from positive to negative. This drives the DNA and electrolyte velocities from positive to negative. Thus, $\mbox{K}^{+}$ ions cancel the DNA mobility inversion by weaking the charge reversal effect induced by ion correlations. 

In the next paragraph, we will discuss the effect of charge reversal on ion currents induced by a pressure gradient.

\subsubsection{Inversion of hydrodynamically induced ion currents through nanopores.}

We now investigate the effect of charge correlations on streaming currents during hydrodynamically-induced polymer translocation events~\cite{buyuk15}. The DNA-membrane geometry including the electrolyte mixture is the same as in the previous section. The only difference is that in hydrodynamically-induced transport experiments, the externally applied voltage difference in Figure 3 is replaced with a pressure gradient $\Delta P_z>0$ at the pore edges. The resulting ionic current thorough the nanopore of length $L=340$ nm and radius $d=3$ nm is given by the number of charges flowing per unit time through the cross section of the channel,
\be\label{eq45}
I_{str}=2\pi e\int_{a^*}^{d^*}drr\rho_c(r)u_s(r).
\ee
In Eq.~(\ref{eq45}), we introduced the effective pore and polymer radii $d^*=d-a_{st}$ and $a^*=a-a_{st}$ where $a_{st}=2$ {\AA} stands for the Stern layer accounting for the stagnant ion layer in the vicinity of the charged pore and DNA surfaces. The charge density function $\rho_c(r)$ is defined by Eq.~(\ref{eq38}). The streaming current velocity is given in turn by the solution of the Stokes equation with an applied pressure gradient,
\be\label{eq46}
\eta\Delta u_s(r)+\frac{\Delta P_z}{L}=0.
\ee
Solving Eq.~(\ref{eq46}), we impose the boundary conditions $u_s(d^*)=0$ (no-slip) and $u_s(a^*)=v$. We also account for the fact that the viscous friction force $F_v=2\pi a^*\eta u_s'(a^*)$ vanishes in the stationary state of the flow. Integrating Eq.~(\ref{eq46}) under these conditions, the streaming flow velocity follows in the form of a Poisseuille profile, 
\be\label{eq47}
u_s(r)=\frac{\Delta P_z}{4\eta L}\left[{d^*}^2-r^2+2{a^*}^2\ln\left(\frac{r}{d^*}\right)\right].
\ee
The knowledge of the charge density~(\ref{eq38}) and liquid velocity~(\ref{eq47}) completes the calculation of the ionic current of Eq.~(\ref{eq45}).
\begin{figure}
\begin{center}
\includegraphics[width=0.7\linewidth]{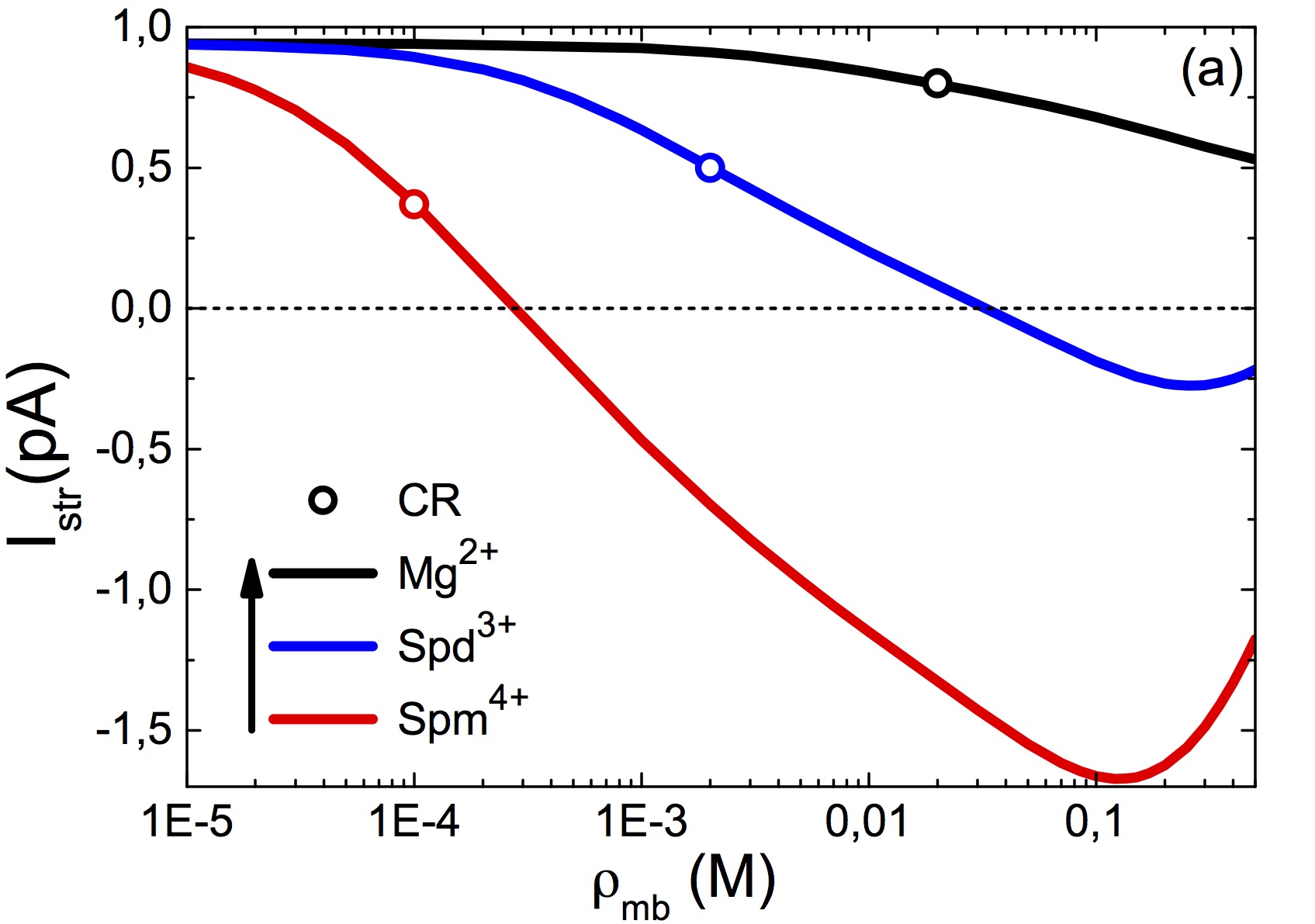}
\includegraphics[width=0.7\linewidth]{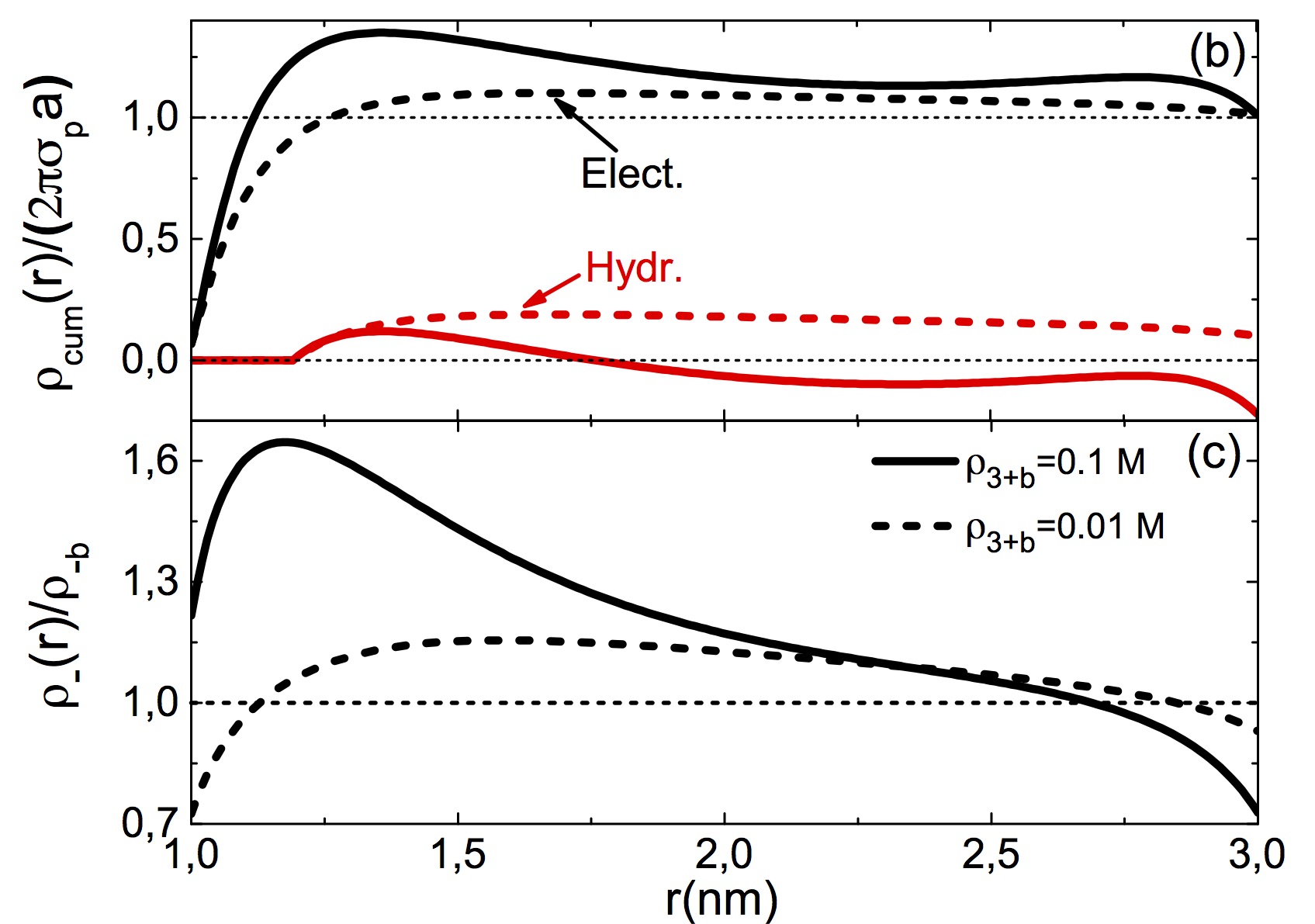}
\caption{(Color online) (a) Streaming current curves at a pressure gradient $\Delta P_z=1$ bar against the reservoir density of the multivalent counter-ion species listed in the legend. Open circles mark the DNA charge reversal (CR) points. (b) Electrostatic (black curves) and hydrodynamic (red curves) cumulative charge densities, and (c) $\mbox{Cl}^-$ densities in the  $\mbox{KCl}+\mbox{SpdCl}_3$ liquid at the reservoir concentrations $\rho_{3+b}=0.01$ M (dashed curves) and $0.1$ M (solid curves). The neutral nanopore ($\sigma_m=0$)   contains a double-stranded DNA molecule of charge density $\sigma_p=0.4$ $e/\mbox{nm}^2$, with a bulk $\mbox{K}^+$ density $\rho_{+b}=0.1$ M in all figures. }
\end{center}
\label{st}
\end{figure}

In Figure 6 (a), we report streaming currents of the electrolyte mixture $\mbox{KCl}+\mbox{ICl}_m$ against the reservoir density of the multivalent cation species $\mbox{I}^{m+}$ in a neutral pore. At weak multivalent ion densities the current is positive. This corresponds to the MF-regime where the negatively charged translocating DNA attracts cations into the pore.  Increasing the bulk magnesium concentration in the $\mbox{KCl}+\mbox{MgCl}_2$ liquid, in agreement with MF ion transport picture, the streaming current drops slightly but remains positive. However, in the liquids containing spermidine and spermine ions, at a characteristic multivalent ion density, the current turns from positive to negative. It is noteworthy that this streaming current reversal has been previously observed in nanofluidic transport experiments through charged nanoslits without DNA~\cite{vdheyden06}. 

The positive ion currents of Figure 6 (a) indicating a net negative charge flow through the pore cannot be explained by MF-transport theory. In order to explain the underlying mechanism behind the current reversal we plot in Figure 6 (b)  the electrostatic cumulative charge density $\rho_{\rm{cum}}(r)=2\pi\int_a^r\mathrm{d}r'r'\rho_{c}(r')$ and the hydrodynamic cumulative charge density  $\rho^*_{\rm{cum}}(r)=2\pi\int_{a^*}^r\mathrm{d}r'r'\rho_{c}(r')$ of the  $\mbox{KCl}+\mbox{SpdCl}_3$ liquid normalized by the DNA charge. The hydrodynamic charge density accounts only for the mobile charges contributing to the streaming flow. Figure 6 (c) displays the chloride densities between the DNA and pore surfaces. At the bulk spermidine concentration $\rho_{3+b}=0.01$ M, Figure 6 (b) shows that the electrostatic cumulative charge density exceeds slightly the DNA charge. This is the sign of a DNA charge reversal effect. This in turn leads to a weak $\mbox{Cl}^-$ excess $\rho_{-}(r)>\rho_{-b}$ between the pore and the DNA (see Figure 6 (c)). However, because the charge reversal and the resulting chloride attraction is weak, the hydrodynamic flow charge dominated by counterions is positive, i.e. $\rho^*_{cum}>0$ for $a^*<r<d^*$. An increase of the spermidine density to $\rho_{3+b}=0.1$ M where one arrives at the inverted current regime in Figure 6 (a), the intensified DNA charge reversal results in a much stronger $\mbox{Cl}^-$ attraction into the pore (see Figures 6 (b) and (c)). This strong anion excess leads in turn to a negative hydrodynamic charge density $\rho_{cum}(d)<0$ and a negative streaming current through the pore. 

To conclude, these calculations show that ionic current inversion during hydrodynamically induced DNA translocation events results from the anion excess in the pore induced in turn by the DNA charge reversal. It is important to note that, similar to the electrophoretic DNA transport of the previous section, the observation of streaming current reversal necessitates a strong DNA charge inversion for the adsorbed anions to bring the dominant contribution to the hydrodynamic flow. This is again illustrated in Figure 6(a) where the charge reversal densities (open circles) are seen to be lower than current inversion densities by several factors. 


\subsection{Solving SC equations in dielectrically inhomogeneous systems}

In the previous paragraphs, we have considered the consequences of the charge reversal effect in different settings. In the next section, we
will focus on the image-charge effects in dielectrically inhomogeneous systems.

\subsubsection{Inversion scheme for the solution of the SC equations.}

In this subsection we present the solution of the SC Eqs.~(\ref{eq10})-(\ref{eq11}) in dielectrically inhomogeneous systems. The technical details of the solution scheme that can 
be found in Refs.~\cite{buyuk14I,buyuk12} will be briefly explained for the case of neutral interfaces. In this case where the average potential $\psi({\bf r})$ is zero, the 
fluctuation-enhanced Poisson-Boltzmann Eq.~(\ref{eq10}) vanishes. This leaves us with Eq.~(\ref{eq11}) to be solved in order to evaluate the ion density
\be
\label{eq48}
\rho_{i}({\bf r})=\rho_{b}e^{-\frac{q^2}{2}\delta v({\bf r})},
\ee
with the ionic self-energy $\delta v(\bf{r})$ defined by Eq.~(\ref{eq12}). The iterative solution strategy consists in recasting the differential equation~(\ref{eq11}) in the form of an 
integral equation. To this aim, we reexpress Eq.~(\ref{eq14}) as
\bea\label{eq49} 
\left[\nabla^2-\kappa_b^2e^{-V_w(\bf{r})}\right]v(\bf{r},\bf{r'})&=&-4\pi\ell_B\delta({\bf r}-{\bf r'}) = \nonumber 
\eea
\bea
-4\pi\ell_B\delta n({\bf r})v({\bf r},{\bf r'})
\eea
where we defined the number-density correction function
\be\label{eq50}
\delta n({\bf r})=2q^2\rho_be^{-V_w(\bf{r})}\left[1-e^{-\frac{q^2}{2}\delta v(\bf{r})}\right].
\ee
Introducing now the DH-kernel
\be\label{eq51}
v_0^{-1}({\bf r},{\bf r'})=-\frac{1}{4\pi\ell_B}\left\{\nabla^2-\kappa_b^2e^{-V_w({\bf r})}\right\}\delta({\bf r}-{\bf r'})
\ee
and using the definition of the Green's function~(\ref{eq16}), one can invert Eq.~(\ref{eq49}) and finally obtain
\be
\label{eq52}
v({\bf r},{\bf r'})=v_0({\bf r},{\bf r'})+\int\mathrm{d}{\bf r''}v_0({\bf r},{\bf r''})\delta n({\bf r''})v({\bf r''},\bf{r'}).
\ee
Eq.~(\ref{eq52}) expresses the solution of the SC-kernel Eq.~(\ref{eq11}) as the sum of the Debye-H\"uckel potential $v_0({\bf r},{\bf r'})$ solution to Eq.~(\ref{eq49}) 
and a correction term associated with the non-uniform charge screening induced by image-charge forces. The iterative solution scheme of Eq.~(\ref{eq52}) consists in replacing at the first iterative step the potential $v({\bf r},{\bf r'})$ on the rhs by the DH potential $v_0({\bf r},{\bf r'})$,  inserting the output potential $v({\bf r},{\bf r'})$ into the rhs of the equation at the next iterative level, and continuing this cycle until numerical convergence is achieved. 
The solution scheme for charged interfaces/nanopores is based on the same inversion idea but technically more involved. The more general scheme can be found in 
Refs.~\cite{buyuk14I,buyuk12}.
\begin{figure}
\begin{center}
\includegraphics[width=0.8\linewidth]{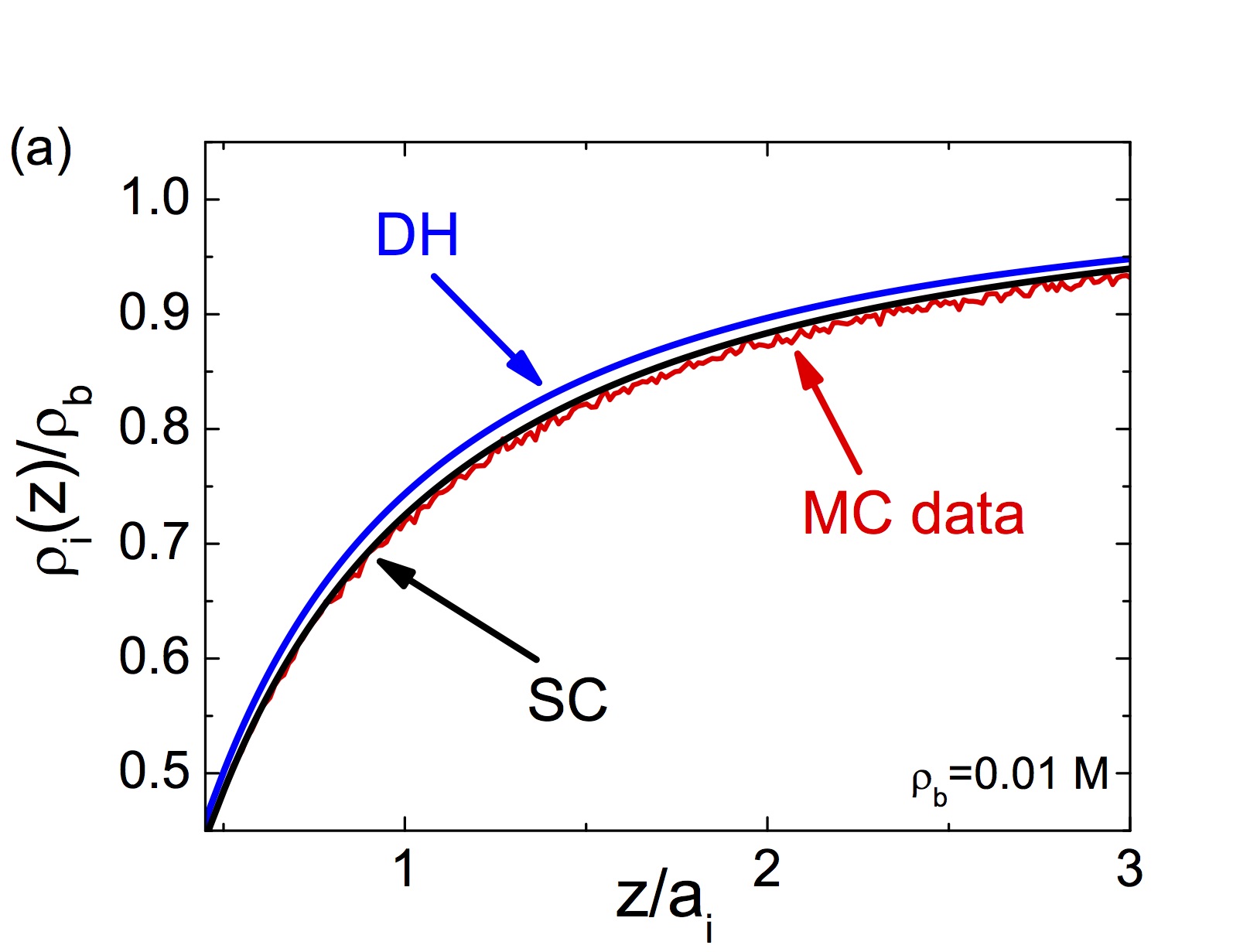}
\includegraphics[width=0.8\linewidth]{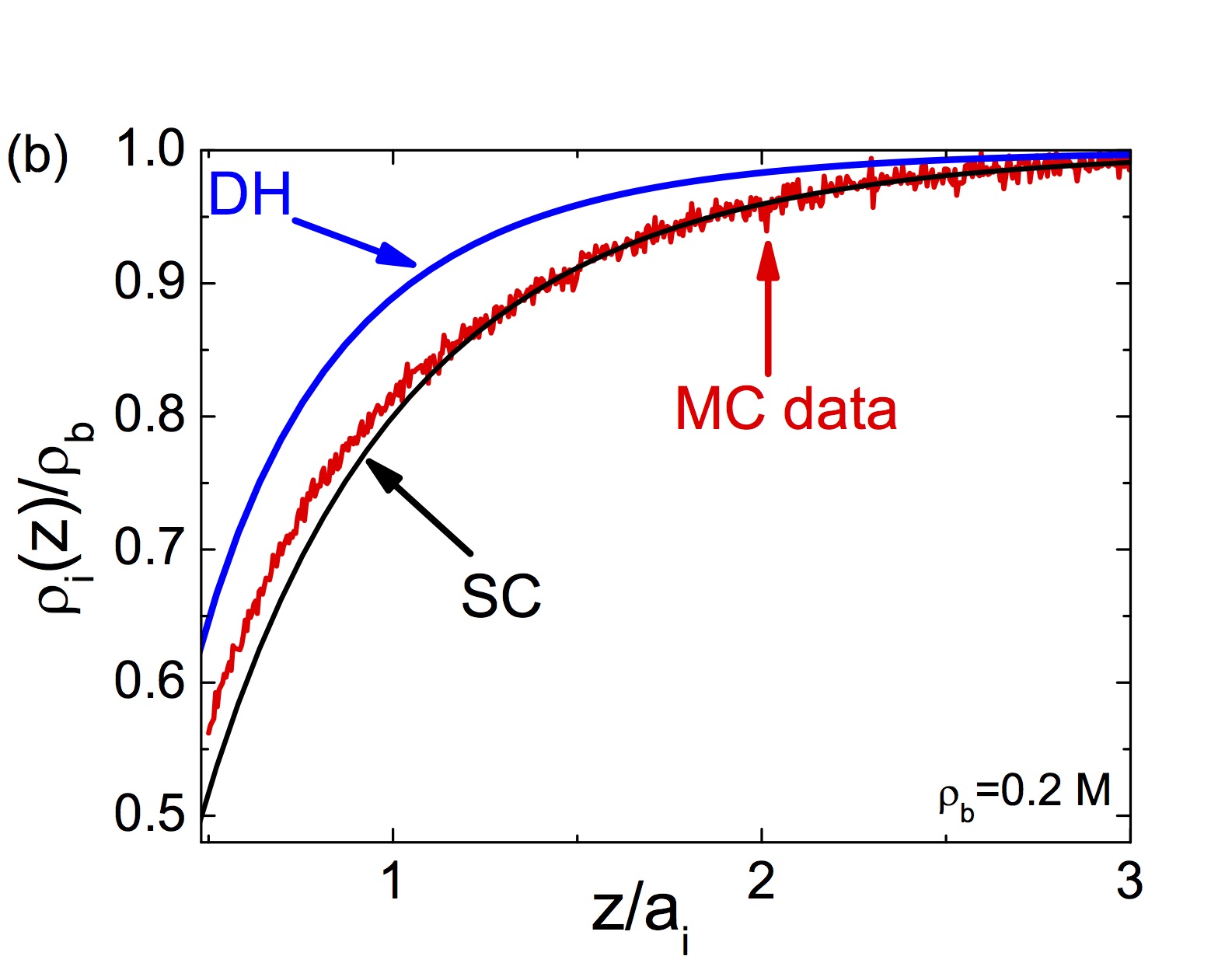}
\caption{(Color online) Ion density profiles at the dielectric interface against the distance from the surface with $\varepsilon_m=1$, $\varepsilon_w=80$, and ion diameter $a_i=4.25$ {\AA} at the bulk ion concentration (a) $\rho_b=0.01$ M and (b) $\rho_b=0.2$ M. The red lines are MC simulation data, the blue lines the DH theory, and the black lines are from the SC scheme~(\ref{eq52}). The theoretical curves and MC data are from Ref.~\cite{buyuk12}.}
\end{center}
\label{Figure7}
\end{figure}

\subsubsection{Image-charge induced correlations at planar interfaces.}

In Figure 7, we display the monovalent ion density profiles obtained from the DH and the SC theories that we compare with MC simulations~\cite{buyuk12}. The dielectric interface located at $z=0$ is neutral and the membrane permittivity is $\varepsilon_m=1$. We emphasize that this configuration is also relevant to the water-air interface whose surface tension was first calculated by Onsager and Samaras~\cite{onsager}. Because their calculation was based on the uniform screening approximation corresponding in our case to the DH approach, the latter is called as well the Onsager approximation. The separation distance is rescaled by the ion size $a_i$, introduced in order to stabilize the MC simulations. At the salt density $\rho_b=0.01$ M (top plot), the SC result exhibits a very good agreement with MC simulations while the DH-theory slightly deviates from the MC data, although the error is minor. At this bulk salt concentration where the electrostatic coupling parameter $\Gamma=\kappa_b\ell_B\approx0.2$ corresponds to the weak-coupling regime, 
the accuracy of the DH-theory is expected. At the much higher salt density $\rho_b=0.2$ M (bottom plot), the SC-theory exhibits a good agreement with MC data but the DH-result overestimates the ion density over the whole interfacial regime. The weak deviation of the SC theory from the MC data is likely to result from excluded volume effects related to ion size but absent in the SC theory. Although the inclusion of ion size is still an open issue, the excluded volume effects can be included into the present SC theory via
repulsive Yukawa interactions (see e.g. Ref. [24]).

The inaccuracy of the DH-result is due to the fact that the ion density $\rho_b=0.2$ M corresponds to the intermediate coupling regime $\Gamma\approx1$. The overestimation of the 
ion density stems from the non-uniform salt screening of the image-charge potential at the interface. The mobile ions that screen this potential being also subject to image-charge forces, the interfacial charge screening is lower than the bulk screening. As the DH-theory assumes a constant screening parameter $\kappa_b$, it cannot account for the reduced screening at the interface. In the SC-theory, this effect is taken into account by the second term on the rhs of Eq.~(\ref{eq52}), correcting the uniform screening approximation of the DH-theory. We also note that in the close vicinity of the interface, the SC-theory slightly deviates from the MC result. This may be due to ionic hard-core effects neglected so far in the SC-formalism.

\subsubsection{Image-charge effects on ion transport through $\alpha$-Hemolysin pores.}

The most significant implication of image-charge correlations are found in electrophoretic charge transport through strongly confined $\alpha$-hemolysin pores. The particularly low conductivity of these pores cannot be explained by the MF electrophoresis. As in the previous section, we will model the pore as a neutral cylinder with radius $d=8.5$ {\AA} and length $L=10$ nm~\cite{clarke09} (see Figure 3). In the most general case, the pore may be blocked by a single-stranded DNA molecule with radius $a=5.0$ {\AA}~\cite{clarke09}. The pore alsocontains the monovalent electrolyte solution KCl. Under an external potential gradient $\Delta V$, the total velocity of the positive or negative ionic species $u_\pm(r)=u_c(r)+u_{T\pm}(r)$ is composed of the convective velocity $u_c(r)$ given by Eq.~(\ref{eq44}), and the drift velocity 
\be
\label{eq53}
u_{T\pm}(r)=\pm\mu_\pm\frac{\Delta V}{L},
\ee
where $\mu_\pm$ stands for the ionic mobility. The ionic current is given by
\be
\label{eq54}
I=2\pi e\sum_{i=\pm}q_i\int_a^d\mathrm{d}rr\rho_i(r)u_i(r).
\ee
Inserting the total velocity $u(r)$ and the ion number density $\rho_i(r)$ into Eq.~(\ref{eq54}), the conductivity takes the form of a linear response relation $I=G\Delta V$, where $G$ stands for the pore conductance (see Ref.~\cite{buyuk14V} for its functional form). 
\begin{figure}
\begin{center}
\includegraphics[width=0.8\linewidth]{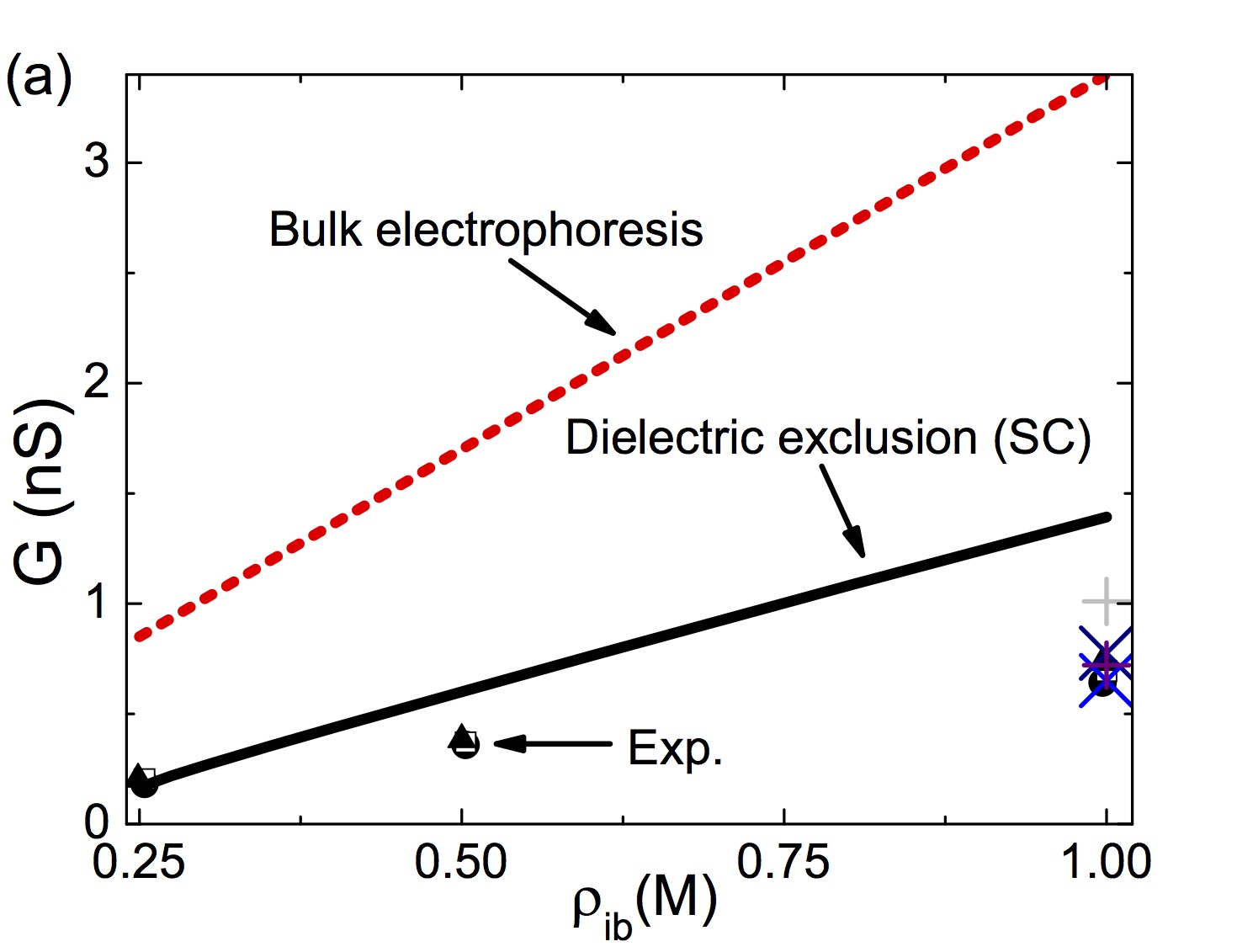}
\includegraphics[width=0.8\linewidth]{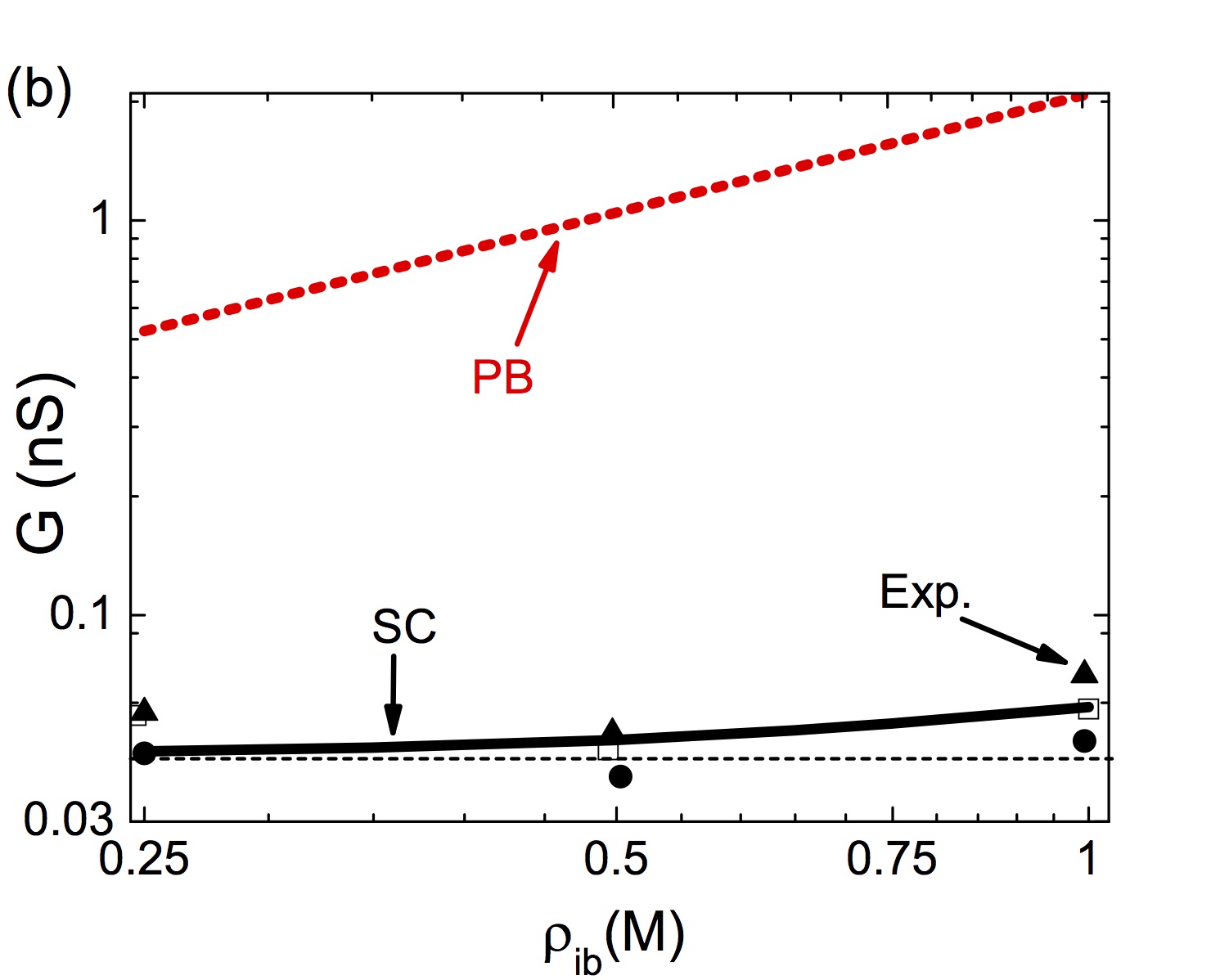}
\caption{(Color online)  (a) Conductivity of an $\alpha$-hemolysin pore against the reservoir concentration of the KCl solution. The pore, modelled as an overall neutral cylinder ($\sigma_m=0.0$ $\mathrm{C/m}^2$), has a radius $d=8.5$ {\AA} and a length $L=10$ nm. Experimental data: black circles, triangles, and open squares from Figure 2 of Ref.~\cite{clarke09}, and additional data at bulk ion density $\rho_{ib}=1.0$ M from Ref.~\cite{miles01} (plus symbol)  and Table 1 of Ref.~\cite{gu00} (cross symbols). (b) The pore in (a) is blocked by a  single-strand DNA molecule with a radius $a=5.0$ {\AA}~\cite{clarke09} and a dielectric permittivity $\varepsilon_p=50$. The effective smeared charge distribution of the ss-DNA is fixed to a value $\sigma_p=0.012$ $\mathrm{e/nm}^2$ providing the best fit with the experimental data (symbols) taken from Figure 3 of Ref.~\cite{clarke09}.}
\end{center}
\label{Figure8}
\end{figure}

In Figure 8 (a), we illustrate the conductance of a DNA-free $\alpha$-hemolysin pore against the reservoir salt density. It is seen that the classical bulk conductivity $G=\pi e\rho_{ib}(\mu_++\mu_-)d^2/L$ overestimates the experimental conductance data by an order of magnitude. However, the SC-theory that can account for image-charge interactions lowers the 
MF-theory to the order of magnitude of experimental data, with a quantitative agreement at low ion densities and a qualitative agreement at large concentrations. The weak pore conductivity is induced by image-charge interactions between the low permittivity membrane and the nanopore. The radius of these pores being comparable with the Bjerrum length 
$d\approx\ell_B$, this results in strongly repulsive polarization forces excluding ions from the pore medium and reducing the net conductance. 

Figure 8 (b) displays the conductance of the same $\alpha$-hemolysin pore blocked now by a single-stranded DNA (see the caption for the characteristic parameters of the DNA molecule). One notes that unlike the conductance of DNA-free pores exhibiting a linear increase with the salt density the blocked pore conductance rises non-linearly at large 
densities but weakly changes with salt at dilute concentrations.  As the PB-conductivity increases linearly with salt density (dashed red curve), the non-linear behaviour of the blocked pore conductivity is clearly a non-MF effect. One also notes that the SC theory can reproduce accurately the non-linear shape of the experimental conductivity data. 

In Ref.~\cite{buyuk14V}, it is shown that the low density conductance of the blocked pore is given by 
\be\label{eq55}
G=\frac{2\pi e}{L}\mu_+\sigma_pa.
\ee
The limiting law~(\ref{eq55}) is displayed in Figure 8 (b) by the dashed horizontal curve. This law is independent of the salt concentration and depends only on the mobility of counterions. Indeed, this counterion-driven charge conductivity results from the competition between repulsive image-charge interactions and attractive DNA-counterion interactions driven by the pore electroneutrality. More precisely, as one lowers the bulk ion density, image charge forces strongly excluding ions cannot lead to a total ionic rejection since a minimum number of ions should stay inside the pore in order to screen the DNA charge. In the dilute concentration regime of Figure 8 (b), these counterions contribute solely to the pore conductance. Hence, the low density limit of the pore conductance corresponds to a non-MF counterion regime whose density is fixed by the DNA surface charge rather than the bulk ion density. At large bulk ion densities, the chemical equilibrium between the pore and bulk media takes over. As a result, coions penetrate into the pore and the conductance starts rising with the bulk salt density. 

This discussion concludes Section 3 of the review, covering the solutions of the self-consistent equations for systems with continuum dielectric properties. In the
subsequent section, we turn to the effects of charge correlations in solvent-explicit electrolyte models.

\section{The dipolar Poisson-Boltzmann equation}

In the previous section we have developed a theoretical treatment of current soft matter problems in which the solvent structure is modeled sufficiently
well by the assumption of continuous dielectric media, in particular of the solvent, and hence can described in terms of a dielectric constant. This description is one of scale:
if one has to consider systems on molecular scales, this assumption becomes questionable. There has therefore been a recent interest to include solvent
properties more explicitly into the Poisson-Boltzmann theory, and one such example is the family of models which include, as a first step, solvent properties
in the form of molecular dipoles. This is done first within the point dipole limit which allows to keep the formulation local. The properties of these theories within
mean-field treatments and their applicationto protein electrostatics have been discussed in a series of papers in the last years 
\cite{coalson96,azuara06,abrashkin07,azuara08,koehl09,koehl10,poitevin11,levy12,levy13,smaoui13}, see also the review \cite{koehl14}.

\subsection{Self-consistent equations for point-dipoles and ions}
\begin{figure}
\begin{center}
\includegraphics[width=0.9\linewidth]{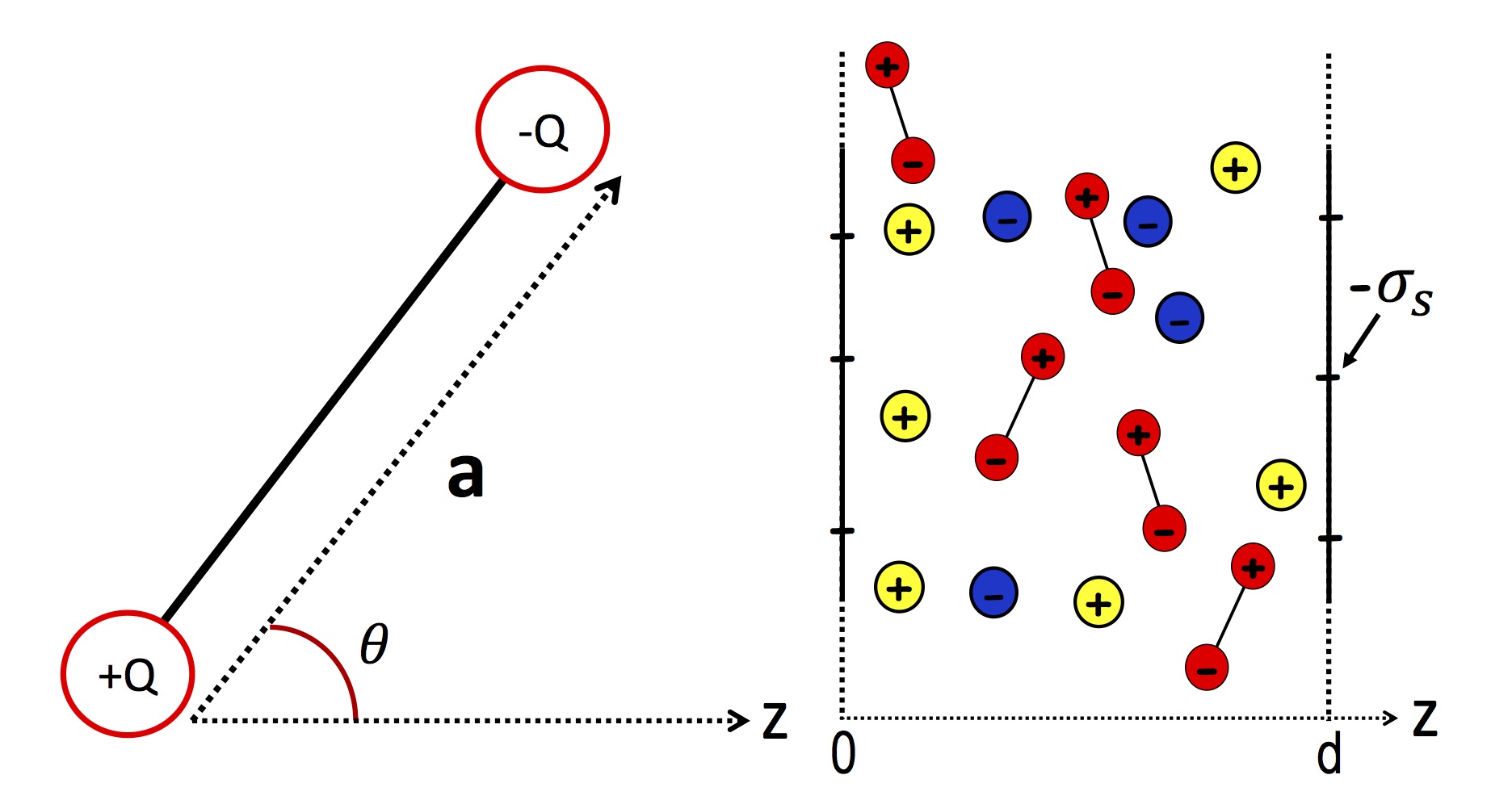}
\caption{(Color online)  Left : Charge geometry of solvent molecules of size $a=1$ {\AA}; the charges of valency $Q=1$ are placed at the ends. Right : 
Geometry of the slit nanopore with surface charge $-\sigma_s\leq 0$ confining solvent molecules (red dipoles), anions (blue circles), and cations (yellow circles).}
\label{Figure9}
\end{center}
\end{figure}
We here derive the variational equations for ions in a point-dipole solvent. The liquid is assumed to be in contact with charged slit membrane walls, though the formulation is 
valid in all geometries. The solvent geometry and the fluid configuration is shown for slit membranes in Figure 7. As the self-consistent equations derived here are 
novel, we present the derivation in detail. 

The grand canonical partition function of ions and dipolar solvent molecules in contact with fixed charge distributions is given by the functional integral over potential configurations $\mathcal{Z}_G=\int \mathcal{D}\phi\;e^{-H[\phi]}$~\cite{coalson96,abrashkin07}, with the Hamiltonian 
\bea\label{eq56}
H[\phi] = \int\mathrm{d}\mathbf{r}\left[\frac{\left[\nabla\phi(\mathbf{r})\right]^2}{8\pi\ell_B(\mathbf{}r)}-i\sigma(\mathbf{r})\phi(\mathbf{r})\right]
\eea
\bea
-\int\frac{\mathrm{d}\mathbf{r}\mathrm{d}\bom}{4\pi}\Lambda_s\;e^{-V_s(\mathbf{r})+i\bp\cdot\nabla\phi} -\sum_i\Lambda_i \int\mathrm{d}\mathbf{r}\;e^{-V_i(\mathbf{r})+iq_i\phi(\mathbf{r})} \nonumber
\eea
where we introduced the dipole moment $\bp=-Q\ba$, and dipolar and ionic wall potentials $V_s(\mathbf{r})$ and $V_i(\mathbf{r})$.  In Eq.~(\ref{eq56}), the first integral term accounts for the electrostatic energy of freely propagating waves in vacuum. It is important to note that the Bjerrum length in vacuum $\ell_B=e^2/(4\pi\varepsilon_0k_BT)$ differs from the one used in the previous chapter for the dielectric continuum model: the dielectric constant of water is absent, as the dipoles explicitly model water properties. The second term of Eq.~(\ref{eq56}) is the contribution from solvent molecules modelled as point dipoles with fugacity $\Lambda_s$. Finally, the third term is due to the mobile ions with fugacity $\Lambda_i$. As stated in the introduction, the SC equations can be derived from a variational extremization principle. The grand potential to be extremized has the form~\cite{netz03}
\be\label{eq57}
\Omega_v=\Omega_0+\lan H-H_0\ran_0,
\ee
where the  bracket $\lan\cdot\ran_0$ denotes the field-theoretic average with respect to the quadratic reference Hamiltonian of the form of Eq.~(\ref{eq2}),
\be\label{eq58}
H_0=\frac{1}{2}\int_{\mathbf{r},\mathbf{r'}}\left[\phi(\mathbf{r})-\psi(\mathbf{r})\right]v^{-1}(\mathbf{r},\mathbf{r'})
\left[\phi(\mathbf{r'})-\psi(\mathbf{r'})\right].
\ee
In Eq.~(\ref{eq58}), we introduced the trial external potential $\psi(\mathbf{r})$ and the electrostatic kernel $v^{-1}(\mathbf{r},\mathbf{r'})$. First, we note that in 
Eq.~(\ref{eq57}), the part of the grand potential corresponding to the Gaussian fluctuations of the electrostatic potential is given by $\Omega_0=-\mathrm{Tr}\ln \left[v\right]/2$. Evaluating  the field-theoretic averages in Eq.~(\ref{eq57}), one obtains the variational grand potential in the form of the following fairly involved expression
\bea\label{eq59}
\Omega_v&=&-\frac{1}{2}\mathrm{Tr}\ln \left[v\right]+\int\frac{\mathrm{d}\mathbf{r}}{8\pi\ell_B}\left[\nabla\psi(\mathbf{r})\right]^2-i\int\mathrm{d}
\mathbf{r}\sigma(\mathbf{r})\psi(\mathbf{r})\nonumber\\
&&+\int\frac{\mathrm{d}\mathbf{r}\;\mathrm{d}\mathbf{r'}}{8\pi\ell_B}\;\delta(\mathbf{r}-\mathbf{r'})\nabla_{\mathbf{r}}\cdot\nabla_{\mathbf{r'}}v(\mathbf{r},\mathbf{r'}) \\
&&-\sum_i\Lambda_i\int\mathrm{d}\mathbf{r}\;e^{-V_i(\mathbf{r})+iq_i\psi(\mathbf{r})-\frac{q_i^2}{2}\int\mathrm{d}\mathbf{r'}\delta(\mathbf{r}-\mathbf{r'})
v(\mathbf{r},\mathbf{r'})} \nonumber
\eea
\bea
-\Lambda_s\int\mathrm{d}\mathbf{r}\frac{\mathrm{d}\bom}{4\pi}\;e^{-V_s(\mathbf{r})+i\bp\cdot\nabla_{\mathbf{r}}\psi(\mathbf{r})
-\frac{1}{2}\int\mathrm{d}\mathbf{r'}\delta(\mathbf{r}-\mathbf{r'})(\bp\cdot\nabla_{\mathbf{r}})(\bp\cdot\nabla_{\mathbf{r'}})v(\mathbf{r},\mathbf{r'})}. \nonumber
\eea
The ionic and solvent number densities follow from the thermodynamic relations 
$\rho_i(\mathbf{r})=\delta\Omega_v/\delta V_i(\mathbf{r})$ and $\rho_s(\mathbf{r})=\delta\Omega_v/\delta V_s(\mathbf{r})$. One finds
\be\label{eq60}
\rho_i(\mathbf{r})=\Lambda_i\;e^{-V_i(\mathbf{r})-q_i\psi(\mathbf{r})-\frac{q_i^2}{2}v(\mathbf{r},\mathbf{r})}
\ee
\be \label{eq61}
\hspace{-4cm}\rho_s(\mathbf{r}) =
\ee
\bea
\Lambda_s\int\frac{\mathrm{d}\bom}{4\pi}\;e^{-V_s(\mathbf{r})+i\bp\cdot\nabla_\mathbf{r}\psi(\mathbf{r})
-\frac{1}{2}\int\mathrm{d}\mathbf{r'}\delta(\mathbf{r}-\mathbf{r'})(\bp\cdot\nabla_\mathbf{r}) (\bp\cdot\nabla_{\mathbf{r'}})v(\mathbf{r},\mathbf{r'})}. \nonumber
\eea
In the bulk region where the electrostatic potential vanishes, $\psi(\mathbf{r})\to 0$, and the propagator satisfies spherical symmetry, 
$v(\mathbf{r},\mathbf{r'})\to v^b(\mathbf{r}-\mathbf{r'})$, one gets from Eqs.~(\ref{eq60})-(\ref{eq61}) the relation between the bulk densities and fugacities as
\bea\label{eq62}
\rho_{ib}&=&\Lambda_i\;e^{-\frac{q_i^2}{2}v^b(\mathbf{r}-\mathbf{r})}\\
\label{eq63}
\rho_{sb}&=&\Lambda_s\;e^{-\frac{1}{2}\int\mathrm{d}\mathbf{r'}\delta(\mathbf{r}-\mathbf{r'})(\bp\cdot\nabla_\mathbf{r})(\bp\cdot\nabla_{\mathbf{r'}})
v^b(\mathbf{r}-\mathbf{r'})}. 
\eea
Passing from Eqs.~(\ref{eq60}),(\ref{eq61}) to Eqs.~(\ref{eq62}),(\ref{eq63}), we accounted for the fact that in the bulk region, the dipolar potential of mean force (PMF) 
in the exponential is independent of the dipolar orientation. 

The SC equations follow from the optimization of the variational grand potential with respect to the electrostatic potential $\psi(\mathbf{r})$ and propagator 
$v(\mathbf{r},\mathbf{r'})$, i.e. $\delta\Omega_v/\delta\psi(\mathbf{r})=0$ and $\delta\Omega_v/\delta v(\mathbf{r},\mathbf{r'})=0$. By evaluating the functional 
derivatives and setting them to zero, replacing the solvent and ion fugacities by the bulk densities according to Eqs.~(\ref{eq62})-(\ref{eq63}), and passing to the 
real electrostatic potential via the transformation $\psi(\mathbf{r})\to i\psi(\mathbf{r})$, one gets after lengthy algebra the variational equations in the form
\bea\label{eq64}
\nabla\cdot\left[\nabla\psi(\mathbf{r})+4\pi\ell_B\bP(\mathbf{r})\right]+4\pi\ell_B\sum_iq_i\rho_i(\mathbf{r}) = \nonumber \\
-4\pi\ell_B\sigma(\mathbf{r})
\eea
\bea
\label{eq65}
\partial_\mu\varepsilon^{\mu\nu}(\mathbf{r})\partial_\nu v(\mathbf{r},\mathbf{r'})-4\pi\ell_B\sum_iq_i^2\rho_i(\mathbf{r})v(\mathbf{r},\mathbf{r'})= \nonumber \\
-4\pi\ell_B\delta(\mathbf{r}-\mathbf{r'}),
\eea 
where we introduced the polarization field
\be\label{eq66}
\bP(\mathbf{r})=-\rho_{sb}\int\frac{\mathrm{d}\bom}{4\pi}\;\bp\;e^{-\varphi_d(\mathbf{r},\bom)}
\ee
that corresponds to the average dipolar polarization, the ion number density
\be\label{eq67}
\rho_i(\mathbf{r})=\rho_{ib}\;e^{-V_i(\mathbf{r})-q_i\psi(\mathbf{r})-\frac{q_i^2}{2}\delta v(\mathbf{r})}\\
\ee
with the renormalized equal-point propagator 
\be\label{eq68}
\delta v(\mathbf{r})=\lim_{\mathbf{r'}\to \mathbf{r}}\left[v(\mathbf{r},\mathbf{r'})-v^b(\mathbf{r}-\mathbf{r'})\right], 
\ee
and the dielectric permittivity tensor
\be\label{eq69}
\varepsilon_{\mu\nu}(\mathbf{r})=\delta_{\mu\nu}+4\pi\ell_B\rho_{sb}\int\frac{\mathrm{d}\bom}{4\pi}\;p_\mu p_\nu\;e^{-\varphi_d(\mathbf{r},\bom)}
\ee
corresponding to the fluctuations of the dipole moment. In Eqs.~(\ref{eq66}) and~(\ref{eq69}), we also introduced the dipolar PMF
\be
\label{eq70}
\varphi_d(\mathbf{r},\bom)=V_s(\mathbf{r})+\bp\cdot\nabla\psi(\mathbf{r})+\varphi_v(\mathbf{r},\bom)
\ee
with the short-hand notation for the part of the dipolar PMF involving the Green's function 
\be\label{eq71}
\varphi_v(\mathbf{r},\bom)=\frac{1}{2}\lim_{\mathbf{r'}\to \mathbf{r}}(\bp\cdot\nabla_\mathbf{r})(\bp\cdot\nabla_{\mathbf{r'}})\left[
v(\mathbf{r},\mathbf{r'})-v^b(\mathbf{r}-\mathbf{r'})\right].
\ee


\subsection{Differential capacitance of low dielectric materials}

\begin{figure}
\begin{center}
\includegraphics[width=0.9\linewidth]{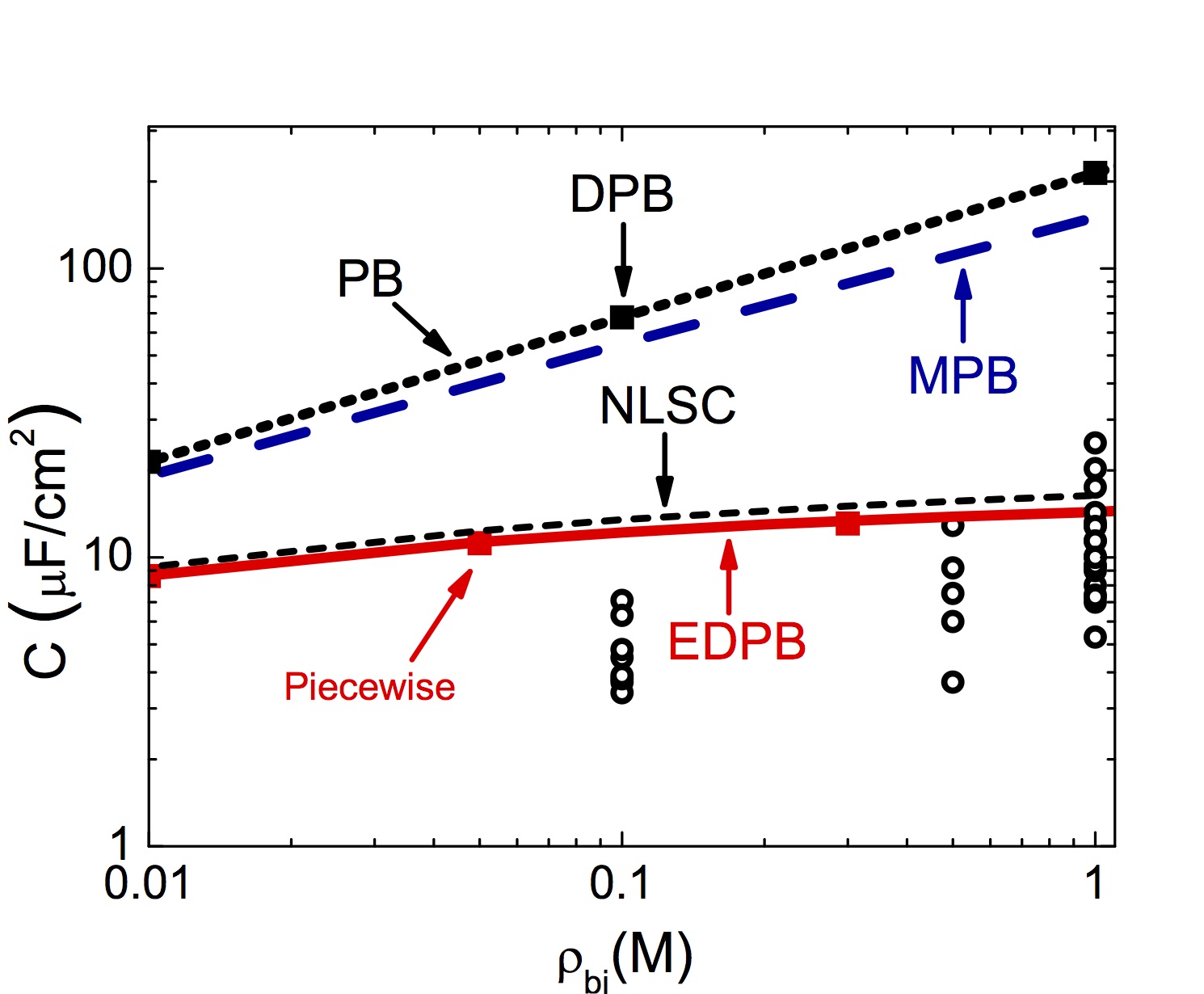}
\caption{(Color online) Differential capacitance against the bulk ion concentration for $\sigma_s\to0$, $\varepsilon_m=1$, and $\varepsilon_w=71$. The black circles are the experimental data, the red solid curve is the result of the EDPB Eq.~(\ref{eq74}), the red squares are from Eq.~(\ref{eq84}), the dashed blue line is the MPB result, the dotted black line is the GC capacitance, and the black squares correspond to the prediction of the DPB equation. The open circles are the experimental capacitance data from \cite{bonthuis11}.}
\end{center}
\label{Figure10}
\end{figure}

The variational equations for the dipolar PB theory being established, we now discuss the effect of the interfacial solvent configuration on the differential capacitance of low dielectric materials such as carbon based substrates of permittivity $\varepsilon_m\approx1$~\cite{buyuk12II}.  We consider a dipolar liquid with bulk density $\rho_{db}=50.8$ M and molecular dipole moment $p_0=1$ {\AA} which corresponds the the bulk dielectric permittivity $\varepsilon_w=71$. The liquid containing a monovalent symmetric electrolyte ($q_+=-q_-=1$) is in contact with a charged dielectric membrane surface located at $z=0$ and separating the dipolar fluid from a membrane with permittivity $\varepsilon_m$. This configuration corresponds to the infinite separation distance limit $d\to\infty$ of the planar geometry of Figure 9. The results obtained in Ref.~\cite{buyuk12II} within a restricted variational ansatz can be rederived from the dipolar SC theory introduced in the previous section. This consists in solving the dipolar equations derived in the previous section by approximating the solution of the kernel Eq.~(\ref{eq65}) with the solution of the DH-equation
\be
\label{eq72}
\nabla^2 v(\mathbf{r},\mathbf{r'})-\kappa_b^2e^{-V_w(\mathbf{r} )-\frac{q^2}{2}\delta v(z)}v(\mathbf{r},\mathbf{r'})=-4\pi\ell_B\delta(\mathbf{r}-\mathbf{r'})
\ee
in planar geometry. For the simple dielectric interface with the liquid located at $z>0$, the solution reads as~\cite{buyuk10}
\be\label{eq73}
v(\mathbf{r},\mathbf{r'})=\frac{\ell_B}{\varepsilon_w}\int\frac{\mathrm{d}^2\bk}{2\pi p_b}e^{i\bk\cdot(\mathbf{r}_\pa-\mathbf{r'}_\pa)}\left[e^{-p|z-z'|}+\Delta_be^{-p(z+z')}\right]
\ee
with the bulk dielectric discontinuity parameter $\Delta_b=(\varepsilon_wp_b-\varepsilon_mk)/(\varepsilon_wp_b+\varepsilon_mk)$. 
Evaluating Eqs.~(\ref{eq67}) and~(\ref{eq68}) with the continuum Green's function~(\ref{eq73}), one gets the fluctuation-enhanced dipolar PB (EDPB) equation~(\ref{eq64}) in the form
\be
\label{eq74}
\frac{\partial}{\partial z}\tilde\varepsilon(z)\frac{\partial\psi(z)}{\partial z}+4\pi\ell_B\sigma(z)+4\pi\ell_B\sum\rho_i(z)q_i=0,
\ee
where we defined the ion number density
\be\label{eq75}
\rho_i(z)=\rho_{b,i}\theta(z)e^{-q_i\psi(z)-V_c(z)}
\ee
with the Heaviside theta function $\theta(z)$, and the local dielectric permittivity function
\be\label{eq76}
\tilde\varepsilon(z)=1+\frac{4\pi}{3}\ell_Bp_0^2\rho_{sb}\theta(z)e^{-U_d(z)}J(z),
\ee
with the ionic and dipolar potentials
\bea
\label{eq77}
V_c(z)&=&\frac{q^2\ell_B}{2\varepsilon_w}\int_0^\infty\frac{\mathrm{d}kk}{\rho_b}\Delta e^{-2\rho_bz},\\
\label{eq78}
U_d(z)&=&\frac{\ell_Bp_0^2}{4\varepsilon_w}\int\frac{\mathrm{d}kk^3}{\rho_b}\Delta e^{-2\rho_bz}\\
\label{eq79}
T_d(z)&=&\frac{\ell_Bp_0^2}{4\varepsilon_w}\int\frac{\mathrm{d}kk}{\rho_c}(2\rho_b^2-k^2)\Delta e^{-2\rho_bz},
\eea
and the auxiliary functions
\bea \label{eq80}
J(z)&=&\frac{3\sqrt\pi}{8T_d^{3/2}(z)}e^{\frac{p_0^2\psi'^2(z)}{4T_d(z)}}\left\{\mathrm{Erf}\left[\Psi_+(z)\right]+ \mathrm{Erf}\left[\Psi_-(z)\right]\right\} \nonumber \\
&&-\frac{3e^{-T_d(z)}}{2T_d(z)}\frac{\sinh\left[p_0\psi'(z)\right]}{p_0\psi'(z)}
\eea
\be \label{eq81}
\Psi_\pm(z)=\frac{2T_d(z)\pm p_0\psi'(z)}{2\sqrt{T_d(z)}}.
\ee
We note that the dielectric permittivity function~(\ref{eq76}) differs from the permittivity tensor~(\ref{eq69}) since the former is related to the average dipole moment~(\ref{eq69}) as $\varepsilon(z)=1+4\pi\ell_B P_z(z)/\psi'(z)$ rather than to dipole moment fluctuations. We also emphasize that in the bulk region, the permittivity tends to the bulk Debye-Langevin form, i.e. $\varepsilon(z\to\infty)=\varepsilon_w=1+4\pi\ell_Bp_0^2\rho_{sb}/3$.
\begin{figure}
\begin{center}
(a)\includegraphics[width=0.7\linewidth]{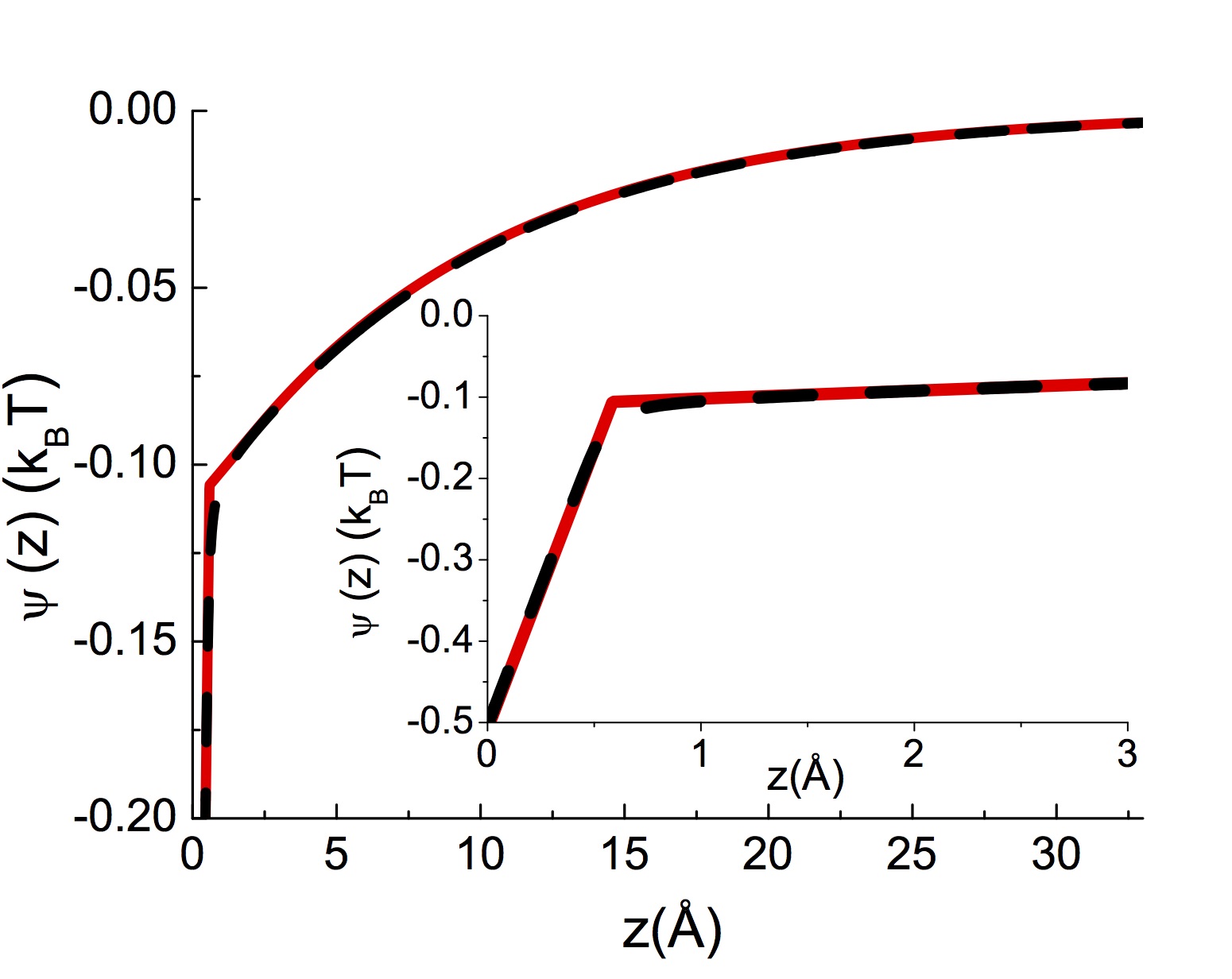}
(b)\includegraphics[width=0.7\linewidth]{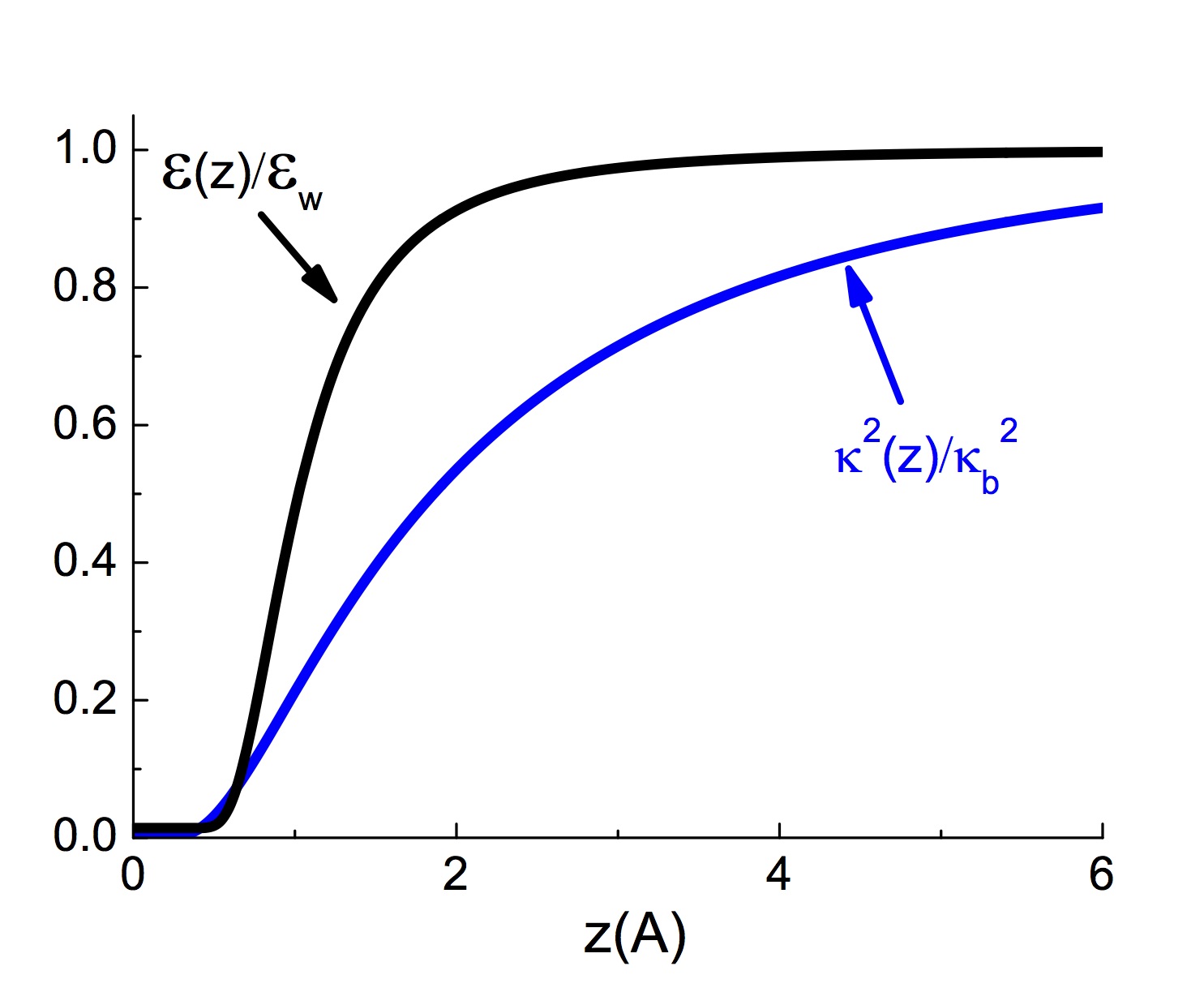}
\caption{(Color online)  (a) Electrostatic potential profile ($\sigma_s=0.01$ $\mbox{nm}^{-2}$) and (b) renormalized density and dielectric permittivity profiles for 
$\varepsilon_w=71$ and $\rho_{bi}=0.1$ M. The red line in (a) is from the restricted variational ansatz Eq.~(\ref{eq83}) and the dashed black line corresponds 
to the solution of the EDPB equation.}
\end{center}
\label{den1}
\end{figure}

The vanishing potential limit of the double layer differential capacitance is given by  
\be\label{eq82}
C_d=\lim_{\sigma_s\to0}\frac{qe^2}{k_BT}\left|\frac{\partial\sigma_s}{\partial\psi(0)}\right|.
\ee
In Figure 10 we plot the capacitance~(\ref{eq82}) from different theories. The open black squares display experimental data obtained for various types of monovalent electrolytes at different densities (see Ref.~\cite{bonthuis11} for details). One sees that both the Gouy-Chapman capacitance (PB) $C_{GC}=(\varepsilon_w\kappa_b)^{-1}$ and the MF
level dipolar PB equation~\cite{coalson96,abrashkin07} overestimate the experimental data by an order of magnitude. The solvent-implicit modified-PB (MBP) equation~\cite{buyuk10} considering exclusively the ion-image interactions slightly lowers the theoretical curve but the disagreement with the experimental data is still very strong. However, the prediction of the EDPB equation~(\ref{eq74}) corrects the PB and DPB results by an order of magnitude and exhibits a good quantitative agreement with the experimental data. 

We further analyze the strong variations of the predictions from one theory to other and to probe the underlying mechanism behind the low capacitance of carbon substrates. Noting that the capacitance~(\ref{eq82}) is directly related to the electrostatic potential, we plotted the latter in Figure 11 (a). We also show in Figure 11 (b) the reduced 
dielectric permittivity profile from Eq.~(\ref{eq76}) and the non-dimensional screening function $\kappa^2(z)/\kappa_b^2=e^{-V_c(z)}$. It is seen that the potential profile is composed 
of three regions (see also the inset). In the vicinity of the charged surface, the potential exhibits a linear drop towards the interface. This strong decrease results from the interfacial dielectric screening deficiency illustrated in Figure 11 (b) by the local permittivity curve. The dielectric permittivity reduction is induced in turn by the interfacial solvent depletion driven by solvent-image interactions and amplifies the PB surface potential by a factor of five. This first layer in the potential profile is followed by a second region where the potential rises linearly but with a lower slope. The linear behaviour of the potential is now due to the charge screening deficiency resulting from image-charge interactions $V_c(z)$ of Eq.~(\ref{eq77}) (see the screening function curve in Figure 11 (b)). Finally, at large separation distances from the interface $z\gtrsim5$ {\AA}, the potential decays exponentially fast. This region corresponds to the diffuse layer governed by charge screening. The latter is the only effect taken into account by the PB-equation. 

In order to have a deeper analytical insight into the behavior of the capacitance data in Figure 10, one can derive an analytical capacitance formula. To this aim, we will  develop a restricted variational approach. Based on the piecewise shape of the electrostatic potential in Figure 11 (b), we assume that the potential $\psi(z)$ is the solution of Eq.~(\ref{eq74}) with a piecewise permittivity profile $\varepsilon(z)=\theta(h_s-z)+\varepsilon_w\theta(z-h_s)$ and screening parameter $\kappa(z)=\kappa_c\theta(z-h_c)$. In these functions, the trial parameters $h_c$ and $h_s$ corresponding respectively to the charge and solvent depletion lengths will be determined from a numerical extremization of the grand potential~(\ref{eq57}). Accounting for the continuity of the potential $\psi(z)$ and the displacement field $\varepsilon(z)\psi'(z)$ at $z=0$, $z=h_s$, and $z=h_c$, the piecewise potential profile takes the form
\bea
\label{eq83}
&& \psi(0<z\leq h_s)= -\frac{2}{\mu\kappa_b}\left[1+\kappa_b(h_c-h_s)\right]+\frac{2\varepsilon_w}{\mu}(z-h_s)\nonumber\\
&& \psi(h_s\leq z\leq h_c)=-\frac{2}{\mu\kappa_b}+\frac{2}{\mu}(z-h_c)\\
&& \psi(z\geq h_c)=-\frac{2}{\mu\kappa_b}e^{-\kappa_b(z-h_c)}.\nonumber
\eea
The trial lengths are determined by inserting the piecewise potential~(\ref{eq83}) into the grand potential~(\ref{eq57}) and optimizing the latter with respect to the parameters $h_s$ and $h_c$. The result of the optimization at bulk ion density $\rho_{ib}=0.1$ M and surface charge $\sigma_s=0.01$ $\mbox{nm}^{-2}$ is shown in Figure 11 (a) by the dashed curve. The comparison with the exact numerical solution of Eq.~(\ref{eq74}) shows that the agreement is very good. 
Evaluating the differential capacitance according to Eq.~(\ref{eq82}) with the piecewise potential~(\ref{eq83}), one finds
\be\label{eq84}
C_d=\frac{\varepsilon_w\kappa_b}{1+\kappa_b(h_c-h_s)+\varepsilon_w\kappa_b h_s}.
\ee
The prediction of this analytical formula reported in Figure 10 (red squares) shows that the formula reproduces the prediction of the EDPB equation very accurately. Now, we note that the inverse capacitance is composed of three terms, $C_d^{-1}=C_{GC}^{-1}+C^{-1}_{c}+C_s^{-1}$. The first contribution of the Gouy-Chapman capacitance corresponding to the diffuse layer was introduced above. The second contribution $C_c^{-1}=(h_c-h_s)\varepsilon_w$ associated with the ionic depletion layer brings the correction that lowers the capacitance curve to the MPB result in Figure 10. Finally, the third term $C_s^{-1}=h_s$ due the solvent depletion responsible for the interfacial dielectric screening deficiency makes the most important contribution to the differential capacitance by lowering the latter to the order of magnitude of the experimental data. Thus, the particularly weak capacitance of low permittivity carbon-based materials is driven by the hydrophobicity of these materials. 

In the next chapter, we relax the present point-dipole approximation and consider non-local electrostatic interactions associated with the extended charge structure of solvent molecules.

\section{Nonlocality in the dipolar Poisson-Boltzmann equation}

The notion of nonlocal electrostatics has already a long history in soft matter. On sufficiently small scales, the dielectric response of a medium becomes
wave-vector dependent, see \cite{kornyshev97,hildebrandt04,maggs06} and references therein. In many descriptions in the literature, this behaviour has been modeled in a phenomenological fashion by invoking simplified dielectric functions. The relevance of this effect has also been shown in recent MD simulations~\cite{bonthuis12,bonthuis13}.
The advantage of our present description of a `structured' solvent is that we can include non-local effects in a direct and natural way, just by relaxing the condition of point dipoles 
in the DPB-model \cite{buyuk13II}; this will be the topic of the last section of this Topical Review.

\subsection{Non-local self-consistent equations}

In this Section, we discuss the dipolar liquid model introduced in Ref.~\cite{buyuk13II} and the associated non-local self-consistent (NLSC) equations derived in Ref.~\cite{buyuk14II}. 
 Thus, we relax the point-dipole approximation and model the solvent molecules as dipoles with finite molecular size $a$ (see Figure 9). The liquid enclosed in a slit pore of size $d$ includes as well $p$ ionic species, each species $i$ with valency $q_i$. The grand-canonical partition function of this liquid is given by the functional integral~$Z_G=\int \mathcal{D}\phi\;e^{-H[\phi]}$, with the Hamiltonian functional
\bea\label{eq85}
H[\phi]&=&\frac{k_BT}{2e^2}\int\mathrm{d}{\bf r}\;\varepsilon_0({\bf r})\left[\nabla\phi({\bf r})\right]^2-i\int\mathrm{d}{\bf r}\sigma({\bf r})\phi({\bf r})\nonumber\\
&&-\Lambda_s\int\frac{\mathrm{d}{\bf r}\mathrm{d}\bom}{4\pi}e^{-V_s({\bf r},\ba)}e^{iQ\left[\phi({\bf r})-\phi({\bf r}+\ba)\right]}\nonumber\\
&&-\sum_i\Lambda_i\int\mathrm{d}{\bf r} e^{-V_i({\bf r})}e^{iq\phi({\bf r})},
\eea
where  $\varepsilon_0(\bf{r})$ accounts for the dielectric permittivity contrast between vacuum and the membrane of permittivities 
$\varepsilon_0$ and $\varepsilon_m$, respectively. The other parameters have 
the same definition as in Eq.~(\ref{eq56}). We also note that if one takes the point-dipole limit $\ba\to0$ while $p_0=-Q\ba$ remains constant, the dipolar term on the rhs of the functional~(\ref{eq85}) converges to the  dipolar contribution of the point-dipole Hamiltonian~(\ref{eq56}).

The variational grand potential is obtained along the same lines as for the point-dipole model of Section 4. One obtains
\bea\label{eq86}
\Omega_v&=&-\frac{1}{2}\mathrm{Tr}\ln\left[v\right]+\int\mathrm{d}\bf{r}\sigma(\bf{r})\psi(\bf{r}) \nonumber\\
&&+\frac{k_BT}{2e^2}\int\mathrm{d}{\bf{r}}\varepsilon_0({\bf r})\left\{\nabla_{\bf{r}}\cdot\nabla_{\bf{r'}} \left.v(\bf{r},\bf{r'})\right|_{\bf{r'}\to\bf{r}}-\left[\nabla\psi(\bf{r})\right]^2\right\}
\nonumber \\
&&-\sum_i\Lambda_i\int\mathrm{d}{\bf r} e^{-V_i(\bf{r})}e^{-q_i\psi(\bf{r})}e^{-\frac{q_i^2}{2}v(\bf{r},\bf{r})} \\
&&-\Lambda_s\int\mathrm{d}{\bf r}\frac{d\bom}{4\pi}e^{-V_s(\bf{r},\ba)}e^{-Q\left[\psi(\bf{r})-\psi(\bf{r}+\ba)\right]}\; e^{-\frac{Q^2}{2}v_d({\bf r},\ba)}.\nonumber
\eea
We will now derive the NLSC equations in the plane geometry of Figure 9. The general form of these equations can be found in Ref.~\cite{buyuk14II}. By evaluating the functional derivatives of the variational grand potential~(\ref{eq86}) with respect to the trial potentials $v({\bf r},{\bf r'})$ and $\psi({\bf r})$, setting the results to zero, and expanding the resulting kernel equation in a Fourier basis according to Eq.~(\ref{eq22}) in order to take advantage of the plane symmetry, after long algebra, the NLSC equations take the form
\bea\label{eq87}
\hspace{-2cm}\frac{k_BT}{e^2}\partial_z\varepsilon_0(z)\partial_z\psi(z) +  \sum_iq_i\rho_i(z) + 
\eea
\bea
2Q\rho_{sb}\int_{a_1(z)}^{a_2(z)}\frac{\mathrm{d}a_z}{2a}\sinh\left[Q\psi(z+a_z)-Q\psi(z)\right]\;e^{-\frac{Q^2}{2}\delta v_d(z,a_z)} \nonumber \\
=-\sigma(z) \nonumber
\eea
\bea \label{eq88}
-\frac{k_BT}{e^2}\left[\partial_z\varepsilon_0(z)\partial_z-\varepsilon_0(z)p^2(z)\right]\tv_0(z,z') + \nonumber 
\eea
\bea
2Q^2\rho_{sb}\int_{a_1(z)}^{a_2(z)}\frac{\mathrm{d}a_z}{2a}\cosh\left[Q\psi(z+a_z)-Q\psi(z)\right]\; \times \nonumber \\
\times e^{-\frac{Q^2}{2}\delta v_d(z,a_z)}\left\{\tv_0(z,z')-\tv_0(z+a_z,z')J_0(k|a_\pa|)\right\} \nonumber \\
= \delta(z-z')\, .
\eea
In Eqs.~(\ref{eq87}),(\ref{eq88}) we defined the ion density
\be
\label{eq89}
\rho_i(z)=\rho_{ib}e^{-q_i\psi_0(z)}e^{-\frac{q_i^2}{2}\delta v_i(z)-V_i(z)},
\ee
and the ionic and dipolar self-energies
\bea
\label{eq90}
&&\delta v_i(z)=v(z,z)-v^b(0)\\
\label{eq91}
&&\delta v_d(z,a_z)=v_d(z,a_z)-2v^b(0)+2v^b(a),
\eea
with the bulk limit of the variational kernel $v^b({\bf r} - {\bf r'})$, and the auxiliary function $p(z)=\sqrt{k^2+\kappa_i^2(z)}$ where the local charge screening function is defined as
\be \label{eq92}
\kappa_i^2(z)=\frac{e^2}{\varepsilon_0(z) k_BT}\sum_iq_i^2\rho_i(z).
\ee
We introduced as well the integral boundaries imposing the impenetrability of the interfaces, $a_1(z)=-\mathrm{min}(a,z)$ and $a_2(z)=\mathrm{min}(a,d-z)$.

The first NLSC Eq.~(\ref{eq87}) is the fluctuation-enhanced dipolar PB equation. In this equation, the dipolar charge density corresponding to the integral term is seen to depend on the value of the electrostatic potential at all points in the slit, resulting in the non-locality of this integro-differential equation. This shows that the consideration of the extended solvent structure directly givers rise to the non-locality of the electrostatic interactions. The equation~(\ref{eq87}) is coupled to the solution of the second NLSC Eq.~(\ref{eq88}) for the electrostatic Green's function. In the next part, we discuss the solution of these equations within different approximations.

\subsection{Solution of NLSC equations at the MF level and beyond}

\begin{figure}
\begin{center}
\includegraphics[width=0.8\linewidth]{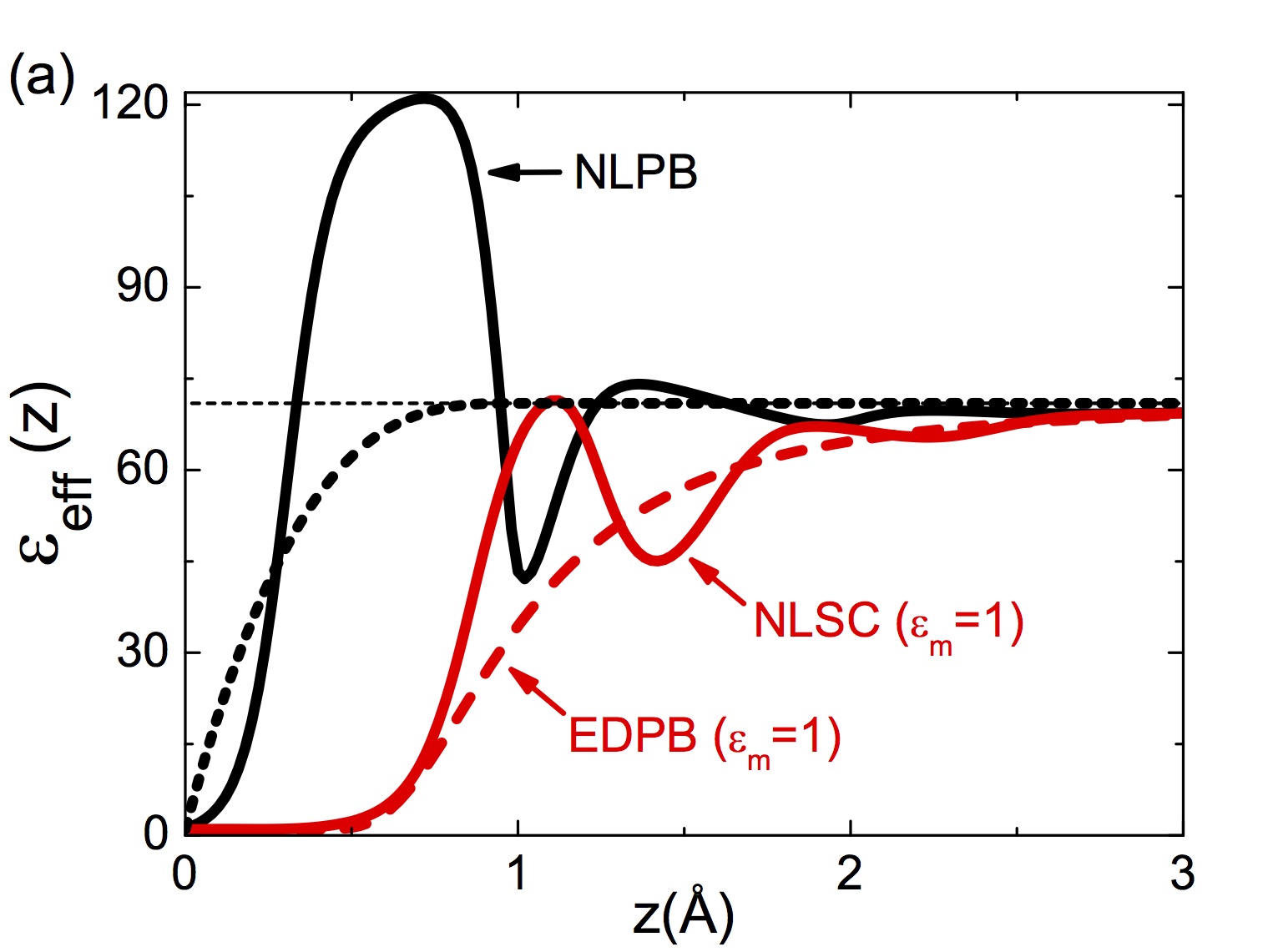}
\includegraphics[width=0.8\linewidth]{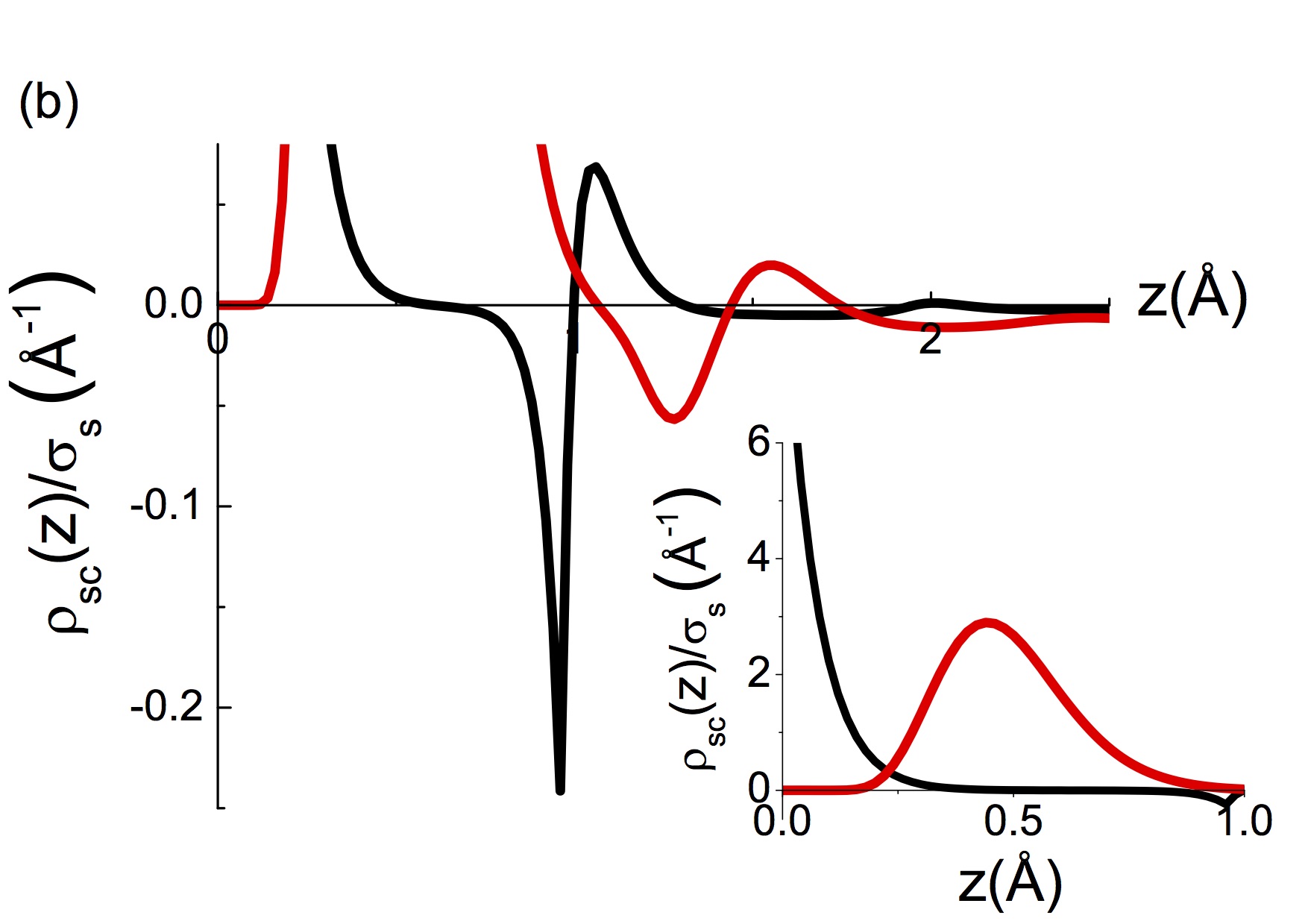}
\caption{(Color online)  (a) Effective dielectric permittivity profile and (b) rescaled polarization charge density obtained from the MF NLPB Eq.~(\ref{eq93}) (solid black curves) and the NLSC Eq.~(\ref{eq87}) (solid red curves) for the solvent with bulk density $\rho_{sb}=50.8$ M and permittivity $\varepsilon_w=71$, in contact with a planar interface with surface charge $\sigma_s=10^{-6}$ e $\mbox{nm}^{-2}$. Salt concentration is $\rho_{ib}=10^{-5}$ M. In (a), the dashed red curve is the permittivity profile of the point-dipole theory (DPB) Eq.~(\ref{eq74}). The dotted black curve displays the dilute solvent limit~(\ref{eq101}) of the NLPB permittivity~(\ref{eq100}).}
\end{center}
\label{figfin}
\end{figure}
We will first consider the NLSC equations~(\ref{eq87}) and~(\ref{eq88}) at the MF level \cite{buyuk14II}. In order to take the MF limit, one can evaluate the saddle-point of the Hamiltonian functional~(\ref{eq86}), i.e. $\delta H[\phi]/\delta\phi({\bf r})|_{\phi=\psi}=0$, or set in Eqs.~(\ref{eq87}),(\ref{eq88})  the electrostatic Green's function accounting for charge correlations to zero, $v({\bf r},{\bf r'})\to0$. The non-local PB (NLPB) equation reads as 
\bea\label{eq93}
&&\partial_z^2\psi(z)+4\pi\ell_B\sigma(z)-8\pi\ell_B\rho_i^b q\sinh\left[q\psi(z)\right]\\
&&+8\pi\ell_BQ\rho_s^b\int_{a_1(z)}^{a_2(z)}\frac{da_z}{2a}\sinh\left[Q\psi(z+a_z)-Q\psi(z)\right]=0,\nonumber
\eea
with the Bjerrum length in air $\ell_B=e^2/(4\pi\varepsilon_0k_BT)\approx55$ nm. 

We will first investigate Eq.~(\ref{eq93}) in the linear response regime of weak surface charges. In order to simplify the presentation, we will also assume that there is a single permeable interface at $z=0$. In this case, one can set $a_1(z)\to-a$ and $a_2(z)\to a$. Finally, linearizing the terms which are non-linear in the potential $\psi(z)$, the linear NLPB equation follows in the form
\be
\label{eq94}
\partial_z^2\psi(z)-\kappa_i^2\psi(z)+\kappa_s^2\int_{-a}^a\frac{da_z}{2a}\left[\psi(z+a_z)-\psi(z)\right] = \nonumber
\ee
\be 
-4\pi\ell_B\sigma(z) 
\ee
with the surface charge distribution $\sigma(z)=-\sigma_s\delta(z)$ and the ionic and solvent screening parameters in the air medium $\kappa_i^2 = 8\pi\ell_B\rho_i^bq_i^2$ and
 $\kappa_s^2 = 8\pi\ell_B\rho_s^bQ^2$, respectively.  One can solve Eq.~(\ref{eq94}) in a Fourier basis and obtain
\be\label{eq95}
\psi(z)=-8\ell_B\sigma_s\int_0^\infty\mathrm{d}k\frac{\cos(kz)}{\kappa_i^2\varepsilon_w+k^2\tilde{\varepsilon}(k)},
\ee
with the Fourier-transformed permittivity 
\be\label{eq96}
\tilde{\varepsilon}(k)=1+4\pi\ell_B\tilde{\chi}(k)
\ee
and the susceptibility function
\be\label{eq97}
\tilde{\chi}(k)=\frac{\kappa_s^2}{4\pi\ell_Bk^2}\left[1-\frac{\sin(ka)}{ka}\right].
\ee
From Eq.~(\ref{eq95}), the electric field $E(z)=\psi'(z)$ follows as
\be\label{eq98}
E(z)=8\ell_B\sigma_s\int_0^\infty\mathrm{d}kk\frac{\sin(kz)}{\kappa_i^2\varepsilon_w+k^2\tilde{\varepsilon}(k)}.
\ee
We note that Eq.~(\ref{eq98}) satisfies the modified Gauss' law $E(z=0)=4\pi\ell_B\sigma_s$, which differs by a factor $\varepsilon_w$ from the usual Gauss's law. 
In the immediate vicinity of the charged interface where $\kappa_iz\ll1$, one can neglect the screening term in the denominator of Eq.~(\ref{eq98}) and rewrite
the electric field as
\be\label{eq99}
E(z)\simeq\frac{4\pi\ell_B\sigma_s}{\varepsilon_{\mathrm{eff}}(z)},
\ee
where we introduced the local effective dielectric permittivity function
\be\label{eq100}
\varepsilon_{\mathrm{eff}}(z)=\frac{\pi}{2}\left/\int_0^\infty\frac{\mathrm{d}k}{k}\frac{\sin(kz)}{\tilde{\varepsilon}(k)}.\right.
\ee
To leading order in the solvent density $O\left((\kappa_sa)^2\right)$, the local permittivity  Eq.~(\ref{eq100}) takes the simple form
\be\label{eq101}
\varepsilon_{\mathrm{eff}}(z)=1+\frac{\left(\kappa_sa\right)^2}{6}\left\{1-\left(1-\frac{z}{a}\right)^3\theta(a-z)\right\},
\ee
where $\theta(z)$ stands for the Heaviside function. Hence, in the case of pure solvents of low permittivity in contact with weakly charged surfaces, the effective permittivity rises from the air permittivity $\varepsilon_{\mathrm{eff}}(0)=1$ to the bulk solvent permittivity $\varepsilon_w=1+\left(\kappa_sa\right)^2/6$. In Fig12 (a), we plotted the non-local permittivity~(\ref{eq100}) (solid black curve) and its dilute solvent limit~(\ref{eq101}) (dotted black curve).  The model parameters are given in the caption. One notes that the behaviour of the non-local permittivity is characterized by fluctuations around the limiting law~(\ref{eq101}), with the first two peaks exceeding the bulk permittivity. We emphasize that this structure formation has been previously observed in molecular dynamics simulations with explicit solvent (see e.g. Ref.~\cite{ballenegger05}) and also in theoretical models based on integral equations~\cite{blum81}. The appearance of these oscillations by the simple consideration of the finite solvent size shows that they are induced by the extended solvent charge structure responsible for the non-local dielectric response of the fluid to the membrane charge. 

In order to trace the microscopic origin of this structure formation, we consider the solvent charge partition close to the membrane surface. To this aim, we will relate the solvent charge density and the effective permittivity in the fluctuation-enhanced limit of the NLSC Eq.~(\ref{eq87}). By integrating this equation from the interface at $z=0$ to any position $z$ in the liquid and using Eq.~(\ref{eq98}), one finds that the dielectric permittivity~(\ref{eq100}) can be expressed as
\be
\label{eq102}
\frac{1}{\varepsilon_{\mathrm{eff}}(z)}=1-\frac{1}{\sigma_s}\int_0^z\mathrm{d}z'\rho_{sc}(z').
\ee
In Eq.~(\ref{eq102}), the solvent charge density is 
\be\label{eq103}
\rho_{sc}(z)=Q\left[\rho_{s+}(z)-\rho_{s-}(z)\right], 
\ee
and the number density of the positive and negative charges making up each solvent molecule is
\be
\label{eq104}
\rho_{s\pm}(z)=\rho_{s}^b\int_{a_1(z)}^{a_2(z)}\frac{\mathrm{d}a_z}{2a}e^{-\frac{Q^2}{2}\delta v_d(z,a_z)}e^{\pm Q\left[\psi(z+a_z)-\psi(z)\right]}.
\ee
The MF-level solvent density is also given by Eq.~(\ref{eq104}) by setting $\delta v_d(z,a_z)=0$. Equation~(\ref{eq102}) is essential as it indicates that the local value of the dielectric permittivity results from the cumulative solvent charge between the interface and the position $z$ in the liquid. We plotted the MF limit of the solvent charge density~(\ref{eq103}) in Figure 12 (b) (black curve). One notes that the solvent response to the charged membrane gives rise to the formation of solvent layers with alternating net charge. More precisely, the immediate vicinity of the interface is characterized by the presence of a positive solvent layer (see the inset) followed by a negative layer (see the main plot) and so on. We also notice that in agreement with Eq.~(\ref{eq102}), the boundary between each solvent layer coincides with the maximum or minimum of the dielectric permittivity profile $\varepsilon_{\mathrm{eff}}(z)$.  Thus, the dielectric permittivity fluctuations are induced by the interfacial solvent charge structure. 

In order to probe dielectric discontinuity effects absent in the MF-NLPB equation, we consider now the NLSC Eqs.~(\ref{eq87}),(\ref{eq88}). Because there exists no recipe for the solution of the non-local kernel Eq.~(\ref{eq88}), we will approximate its solution by the solution of the DH equation of the local dipolar model. This leaves us with the single non-local Eq.~(\ref{eq87}) for the average potential $\psi(z)$ whose numerical solution is also considerably technical. The numerical relaxation schemes that we have developed can be found in Refs.~\cite{buyuk14b} and~\cite{buyuk15II}. In Figures 12 (a) and (b), we plotted the effective permittivity~(\ref{eq102}) and the solvent charge density~(\ref{eq103}) (solid red curves) obtained from the NLSC Eq.~(\ref{eq87}) at the membrane permittivity $\varepsilon_m=1$. In Figure 12 (b), one sees that image-dipole effects repelling solvent molecules from the surface reduces the interfacial solvent charge density. Figure 12 (a)  shows that this result in an amplification of the surface dielectric void layer and an overall reduction of the dielectric permittivity of the MF-NLPB theory. 

The image-dipole correlations amplify the interfacial dielectric void resulting at the MF-level from the non-local solvent response. We now wish to identify the effect of the extended solvent structure beyond MF-level. To this aim, in Figure 12, we compare the dielectric permittivity of the NLSC formalism with the permittivity profile of the point-dipole (EDPB) theory (dashed red curve). One notes that both permittivities are characterized by the same dielectric void layer, with the NLSC permittivity exhibiting fluctuations around the EDP permittivity. This indicates that the strong interfacial dielectric void is driven by solvent-image interactions present both in the local and the non-local formalism. This point is confirmed by Figure 10 where we compare the local and non-local capacitance predictions. It is seen that the NLPB capacitance weakly differs from the EDP result. Since we showed that the low differential capacitance is induced by the surface solvent depletion, the closeness of the local and non-local capacitances is indeed expected. Thus, we can conclude that the non-locality plays no major role in the hydrophobicity of the membrane surface.  
\\

\section{Discussion and Conclusions}

In this topical review we have presented an introduction to the recently developed variational or self-consistent approach to soft matter electrostatics which 
allows to include correlation effects beyond Poisson-Boltzmann theory. We developed two cases: first, the continuum approach in which solvent properties
are captured by its macroscopic dielectric constant, and second, a model with explicit solvent in which the solvent properties are modeled with
point or finite-size dipoles. The latter theory is, to our knowledge, the first explicitly formulated example of a `structured Coulomb fluid' treated beyond
mean-field theory, and including non-local effects.

In this work, we have selected a few application examples and developed them in some technical detail with the intention to help the interested reader 
in gaining access to the approach. We hope that the illustrative examples we discussed do show the power of the method, which nevertheless comes at a certain 
computational effort. In many further applications, the equations will in fact only be solvable by numerical methods which are beginning to be developed.  
The further close confrontation of theoretical results that can be obtained from the self-consistent approach with experimental results and computer simulations 
on soft matter systems will  be crucial for the future development of a systematic approach to correlation effects in soft condensed matter.
\\

{\bf Acknowledgement.} RB gratefully acknowledges support from the ANR-Blanc project {\it Fluctuations in Structured Coulomb Fluids} (FSCF, ANR-12-BSV5-0009)
and FRABio. We thank Ralf Everaers, Anthony Maggs and Henri Orland for discussions.   
\\

\end{document}